\documentclass{emulateapj}
\usepackage{lscape}
\newcommand{\km}{${\rm km\,s}^{-1}$}
\newcommand{\fuse}{{\em FUSE}}
\newcommand{\hi}{H$\;${\small\rm I}\relax}
\newcommand{\hii}{H$\;${\small\rm II}\relax}
 
\newcommand{\ari}{Ar$\;${\small\rm I}\relax}

\newcommand{\civ}{C$\;${\small\rm IV}\relax}
\newcommand{\nni}{N$\;${\small\rm I}\relax}
\newcommand{\nii}{N$\;${\small\rm II}\relax}

\newcommand{\oi}{O$\;${\small\rm I}\relax}

\newcommand{\oiii}{O$\;${\small\rm III}\relax}

\newcommand{\ovi}{O$\;${\small\rm VI}\relax}

\newcommand{\pii}{P$\;${\small\rm II}\relax}

\newcommand{\sii}{S$\;${\small\rm II}\relax}
\newcommand{\siii}{Si$\;${\small\rm II}\relax}
\newcommand{\siiii}{Si$\;${\small\rm III}\relax}

\newcommand{\Siii}{S$\;${\small\rm III}\relax}
\newcommand{\siiv}{Si$\;${\small\rm IV}\relax}

\newcommand{\feii}{Fe$\;${\small\rm II}\relax}
\newcommand{\feiii}{Fe$\;${\small\rm III}\relax}

\newcommand{\D}{$^\circ$}

\slugcomment{Accepted for publication in the ApJ}
\shortauthors{Lehner, Staveley-Smith, \& Howk}
\shorttitle{Properties and Origin of the HVC toward the LMC}
\begin{document}

\title{Properties and Origin of the High-Velocity Gas Toward the  Large Magellanic Cloud\altaffilmark{1}}
\author{N.\ Lehner\altaffilmark{2},
	L. \ Staveley-Smith\altaffilmark{3},
	J.C. \ Howk\altaffilmark{2}
	}
   
\altaffiltext{1}{Based on observations made with the NASA-CNES-CSA Far Ultraviolet Spectroscopic Explorer. FUSE is operated for NASA by the Johns Hopkins University under NASA contract NAS5-32985. Based on observations made with the NASA/ESA Hubble Space Telescope, obtained at the Space Telescope Science Institute, which is operated by the Association of Universities for Research in Astronomy, Inc. under NASA contract No. NAS5-26555. Based on observations made by the Parkes telescope. The Parkes telescope is part of the Autralia Telescope, which is funded by the Commonwealth of Australia for operation as a National Facility managed by CSIRO.
}
\altaffiltext{2}{Department of Physics, University of Notre Dame, 225 Nieuwland Science Hall, Notre Dame, IN 46556}
\altaffiltext{3}{School of Physics M013, University of Western Australia, 35 Stirling Highway, Crawley, WA 6009, Australia }

\begin{abstract}
In the spectra of 139 early-type Large Magellanic Cloud (LMC) stars  observed with \fuse\ and with deep radio Parkes \hi\ 21-cm observations along those stars, we search for and analyze the absorption and emission from high-velocity gas at $+90 \le v_{\rm LSR} \le +175$ \km.  The \hi\ column density of the high-velocity clouds (HVCs) along these sightlines ranges from $<10^{18.4}$ to $10^{19.2}$ cm$^{-2}$. The incidence of the HVC metal absorption is 70\%, significantly higher than the \hi\ emission occurrence of 32\%. We find that the mean  metallicity of the HVC is $[$\oi/\hi$] = -0.51 \,^{+0.12}_{-0.16}$.   There is no strong evidence for a large variation in the HVC metallicity, implying that these HVCs have a similar origin and are part of the same complex. The mean and scatter of the HVC  metallicities  are more consistent with the present-day LMC oxygen abundance than that of the Small Magellanic Cloud or the Milky Way. We find  that on average $[$\siii/\oi$] = +0.48 \,^{+0.15}_{-0.25}$ and $[$\feii/\oi$] = +0.33 \,^{+0.14}_{-0.21}$, implying that the HVC complex is dominantly ionized. The HVC complex has a multiphase structure with a neutral (\oi, \feii), weakly ionized  (\feii, \nii), and highly ionized (\ovi) components, and has evidence of dust but no molecules. All the observed properties of the HVC  can be explained by an energetic outflow from the LMC. This is the first example of a large ($> 10^{6}$ M$_\odot$) HVC complex that is linked to stellar feedback occurring in a dwarf spiral galaxy.
\end{abstract}

\section{Introduction}
The interactions of galaxies and the nearby intergalactic medium (IGM) through the accretion of matter onto galaxies or the expulsion of matter and energy in winds from galaxies are crucial for the evolution of both the galaxies and the IGM. The star formation histories, gas content, and metallicity of a galaxy are co-dependent on both internal processes and on the interaction between the galaxy and the local IGM.  Today's star-forming galaxies are continuing to form by accreting gas from the nearby IGM, from matter condensing out of a hot corona \citep[e.g.,][]{peek08}, from the stripped ISM of smaller dwarf systems \citep[e.g.,][]{putman98}, or even from matter ejected by earlier star formation episodes \citep{bertone07,bouche07}; much of the accretion may proceed through analogs to the high-velocity clouds (HVCs) found about the Milky Way \citep[MW, e.g.,][]{wakker07,lockman08,thom08}.  The continued formation of stars in galaxies without depleting the available gas and the observed metallicity distribution of long-lived stars in our galaxies require a continuous infall of low-metallicity matter \citep[e.g.,][]{vandenbergh62,tinsley81,matteucci03,bland08}. Balancing the accretion of new matter, feedback from strong star formation can drive large-scale circulations of gas away from the disk or outflows that feed matter into the nearby IGM \citep[e.g.,][]{veilleux05,simcoe06,oppenheimer08}. These competing processes are at the heart of the evolution of galaxies and the IGM.  

Most of the current observational information of galactic interaction and winds comes from the observation of emission lines of hydrogen and metals \citep[e.g.,][]{martin98,veilleux05,strickland07}. These observations have  yielded crucial information, such as the multiphase nature (from very hot ionized gas to cold neutral gas) and large-scale morphology of these features. But because their density-squared dependence, emission line observations are heavily biased toward the highest-density regions, which may only trace a relatively small fraction of the total mass and energy \citep{strickland00,veilleux05}. Measurements that rely on absorption lines are less biased to the highest densities, but they  generally probe only directions toward the brightest stellar clusters, losing detail as the stars are integrated in the spectrograph slit, and are restricted to starburst galaxies \citep{heckman01,cannon05}.  However, one counter-example is the Magellanic Clouds that are near enough to allow absorption line measurements toward many individual stars, and thus are excellent laboratories for studying  recycling processes of energy and matter  in low mass, low metallicity galaxies. 

The Large Magellanic Cloud (LMC) is the brightest dwarf disk galaxy in the sky situated at $\sim$50 kpc, while the SMC is an irregular dwarf galaxy a little farther away at $\sim$60 kpc. With a lower metallicity and a higher gaseous mass, the SMC is less evolved than the LMC and has a different star formation history. These galaxies are also much smaller than the MW but are quite common in the Universe, hence they can be used to better understand their impact on their external environments. It has been known for a long time that these galaxies are associated with large gaseous structures between them  (the Magellanic Bridge) and trailing (the Magellanic Stream) or leading (the Leading Arm) them \citep[e.g.,][]{mathewson74,putman98,bruns05}. In addition, many FUV spectra of LMC, SMC, and Magellanic Bridge stars have revealed neutral, weakly and highly ionized at blue-shifted high velocities relative to the Clouds and inconsistent with those of the MW or the Clouds \citep[e.g.,][]{savage81,welty99,lehner01a,danforth02,hoopes02,howk02,lehner02}. These components are signatures of high-velocity clouds (HVC) at $v_{\rm LSR} \sim \,$80--160 \km, previously unseen in \hi\ 21-cm emission observations. Subsequent deep observations toward the LMC have also revealed some of these HVCs in \hi\ 21-cm emission \citep{staveley-smith03}.

From the mid-90's to 2006, one of the favored models predicted that the Magellanic Stream and Leading Arm had formed from a close interaction between the Clouds and the MW about 1.5 Gyr ago where the gas was pulled  from the SMC via tidal forces, while the Bridge was believed to have formed through a close encounter between the LMC and SMC some 200 Myr ago where the gas and stars were pulled from the SMC via tidal interaction \citep{gardiner96}. These models had already some shortcoming as they could not explain  the low metallicity of the Bridge gas and stars \citep{rolleston99,lehner08} or the absence of stars in the Stream. Even more dramatically, the {\em Hubble Space Telescope }\  ({\em HST}) measurements of the proper motions of the LMC and SMC have changed our view of these galaxies and their interaction with the Milky  Way \citep{kallivayalil06,kallivayalil06b,piatek08}. Instead of being long time companions of the MW, these galaxies may be just passing through for the first time, and the LMC/SMC may not be bound to the MW anymore, although, according to \citet{shattow09}, there appears to be still enough uncertainties in the proper motions and circular velocities of the LMC and MW to allow these two galaxies to be bound. Nevertheless these new measurements appear to rule out models that involve mainly tidal forces as these forces are ineffective without multiple passages \citep{besla07,ruzicka09}. Ram pressure from a low density, hot ionized, and extended MW halo may provide the mean to remove a large amount of gas from the disk of the LMC \citep{mastropietro05,mastropietro08}. Alternatively or additionally to ram pressure, stellar feedback from the LMC could provide another source  of gas for the Magellanic Stream \citep{olano04,nidever08,besla07}. But is there any evidence for a generalized stellar feedback beyond the thick disk of the LMC? 

Some evidence was presented that the HVCs at $v_{\rm LSR} \sim \,$100--160 \km\ in front the LMC  may be signatures of a generalized galactic outflow from the LMC \citep{staveley-smith03,lehner07} rather than originating from the MW \citep{deboer90}. Shedding light on the origin of these HVCs is important for our general understanding of galaxy evolution, but appears even more pressing in the context of the new Clouds' proper motions and their implications. We have therefore embarked in an effort to characterize the HVCs\footnote{At $v_{\rm LSR} \sim \,$90--160 \km, most of these clouds are HVCs relative to both the MW and the LMC.} that are between the MW and the Clouds. An unprecedented number of background targets is available behind the HVCs since over the last eight years the {\it Far Ultraviolet Spectroscopic Explorer}\ ({\fuse}) has collected over 230 spectra of early-type stars in the SMC and LMC. In contrast the best studied HVC complex in the UV is complex C with only 11 sightlines  \citep[e.g.,][]{collins07}. Because of  the multiphase nature of these features, the FUV spectrum is ideal to undertake such a study: (1) the FUV spectra are at high enough spectral resolution to estimate the column density and decipher high-velocity absorption relative to the MW or the Clouds, and (2) the FUV bandpass provides access to a wide range of gas phases, from the molecular clouds, to the neutral atoms, to low-, intermediate-, and highly-ionized gas. 

In this paper, we focus on the HVC toward the LMC, with three main aims: (1) to derive the metallicity of these HVCs,  (2) to characterize the physical and ionization conditions, and (3) to describe the distribution of the properties (kinematics, ionization, etc...) of the HVCs. As the distance of these clouds cannot be bracketed to better than $d<50 $ kpc, the LMC distance, the metallicity is one of the most promising ways to determine the origin(s) of the HVCs \citep{savage81,wakker01}  knowing the metallicity of the LMC, SMC, and MW and their chemical evolution \citep[e.g.,][]{russell92,pagel98}.  

In order to estimate the metallicity, we compare the column density of \oi\ and \hi. \oi\ is the best metal proxy for \hi\ since its ionization potential and charge-exchange reactions with hydrogen ensure that the ionization of \hi\ and \oi\  are strongly coupled in galactic environments \citep{jenkins00,lehner03}. Furthermore, oxygen is only  mildly depleted into dust grains \citep[ $\la -0.1$ dex based on the interstellar \oi\ estimate in the galactic disk, e.g.,][]{meyer98,jensen05}, implying that the \oi/\hi\ ratio is the best indicator of the metallicity, especially in partially ionized gas. The ionization level of the gas can be probed with the ratio \feii/\oi. Indeed while Fe can be depleted into dust (hence the Fe/O is a priori expected to be subsolar), we show that this ratio is supersolar, strongly suggesting that the HVC gas is largely ionized (independently confirmed by the systematic presence of \nii\ and \ovi\ absorption). 

Our paper is organized as follows. In \S\ref{sec-data} we describe our sample, the instruments, and methods to derive the velocities and column densities for \oi, \feii, and \hi\ associated with the HVCs, as well as discuss the issues when comparing emission with absorption data  (\oi\ and \feii\ are seen in absorption in the FUV spectra of stars, while \hi\ is obtained from deep \hi\ 21-cm emission observations). The metallicity and relative-abundance estimates of the HVCs are presented in \S\ref{sec-abund}, while in \S\ref{sec-prop} we present their properties (abundances, ionization, depletion, kinematics) and distribution. In \S\ref{sec-disc} and \S\ref{sec-conc} we discuss the implication of our findings and summarize our results, respectively.

\begin{figure*}[!t]
\epsscale{1} 
\plotone{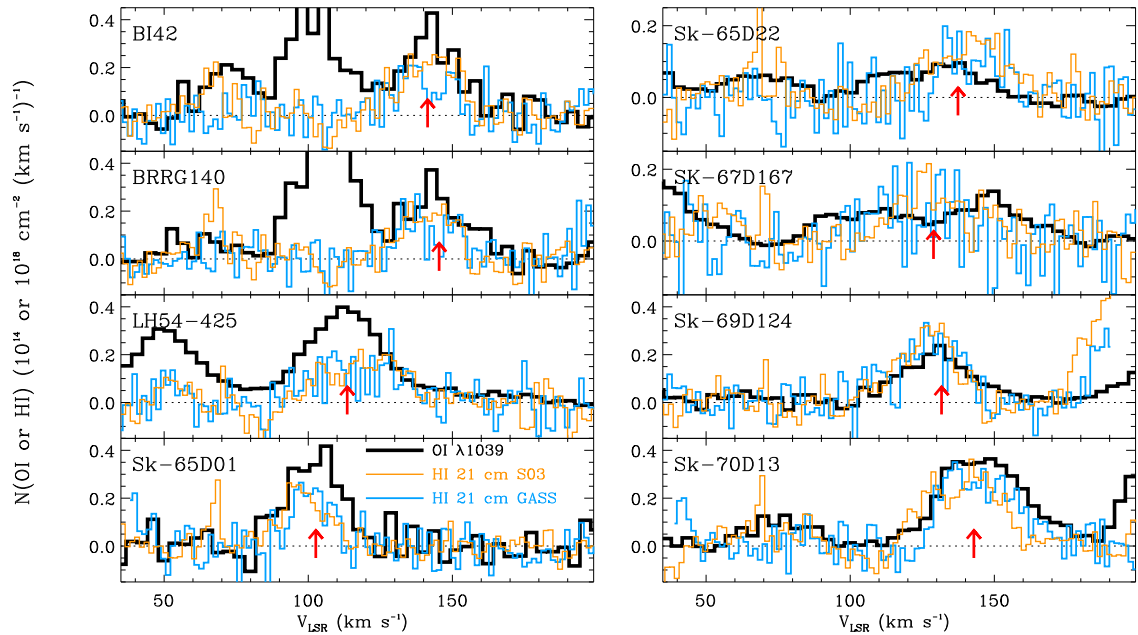}
\caption{Examples of apparent column density profiles for \oi\ (black histogram) and column density  profiles (from the brightness temperature) for \hi\ (blue [GASS] and orange [S03] histograms). The red arrow shows the value of the average velocity of the HVC. The strong feature at 100--110 \km\ in the panels of BI42 and BRRG140 is the contaminant LMC component of 5--0 R(2) H$_2$ at 1038.689 \AA.  Intermediate-velocity components (40--80 \km) or multiple HVC components can also be seen in the profiles.  \label{f-aod}}
\end{figure*}

\section{Data, Analysis, and Results}\label{sec-data}
\subsection{The LMC Sample}

Our first criterion to select the early-type stars for our LMC sample was that their \fuse\ (see next for description of the instrument itself) spectra were adequate for our primary purposes (i.e. to detect and measure the high-velocity \oi\ and \feii\ absorption). We therefore retrieved the 185 fully calibrated spectra from the {\fuse}\ Magellanic Clouds Legacy Project archive at the Multimission Archive at STScI \citep[MAST,][]{blair09}.\footnote{http://archive.stsci.edu/prepds/fuse\_mc/. We also refer the reader to \citet{danforth02} for the first \fuse\ atlas of the LMC, where 57 LMC stars were included with some basic information about the HVC studied here.} For each stars, we created stacks of velocity profiles for key atomic and ionic species  (e.g., \oi, \nni, \ari, \feii, etc...) and several H$_2$ transitions as well as an overall \fuse\ spectrum of the stars. Forty-six stars were rejected from our final sample because either the signal-to-noise in their spectra was too poor or because the H$_2$ absorption was too strong or owing to a possible stellar contamination (including our inability to find a satisfactory continuum near the \oi\ lines), or a combination of these possibilities. In Table~\ref{t-sum}, the first three columns give the name, right ascension  (RA) and declination (DEC) of the 139 stars included in our sample. Among these 139 stars, there are 78 stars where \oi\ $\lambda$1039 high-velocity absorption is detected and  93 stars where \feii\ $\lambda$1144 high-velocity absorption is detected.

\subsection{UV Data and Measurements}
\subsubsection{Instruments and Calibrations}
The available resonance lines of \oi\ and \feii\ are all present in the FUV bandpass observed with \fuse\ and {\em HST}/STIS. Descriptions of the \fuse\ instrument  design and  in-flight performance are found in \citet{moos00} and \citet{sahnow00}, and see also \citet{dixon07} for an update. Information about STIS and its in-flight performance is given in the STIS Instrument Handbook \citep{kim07}. The spectra of ten stars in our sample were obtained with both \fuse\ and STIS E140M (see Tables~\ref{t-sum} and \ref{t-rel}). The spectral resolution for the STIS E140M data is 7 \km\ with a detector pixel size of 3.5 \km, while the \fuse\ resolution is about 20 \km. We note that generally the \fuse\ spectra were obtained through the large aperture (LWRS: $30 \arcsec \times 30 \arcsec$) to maximize throughput, except in very crowded regions where the medium aperture (MDRS: $4\arcsec \times 20\arcsec$) was used to ensure that only the selected star was in the \fuse\ aperture. 

The STIS E140M data processing is fully described in \citet{lehner07} for 4 stars. We applied the same data reduction for the remaining 6 stars and we refer the reader to this paper for further calibration information. The {\fuse}\ data were retrieved fully calibrated from the {\fuse}\ Magellanic Clouds Legacy Project archive. In order to maximize the signal-to-noise (S/N), we use the fully co-added spectra produced by  CalFUSE v3.2 \citep{dixon07}. We systematically crosschecked in the regions of interest that no spurious features contaminated the lines under investigation and that the resolution of the lines was not degraded. The S/N ratios per resolution element are typically $\ga 5$--40 in the \fuse\ spectra.

The absolute \fuse\ wavelength scale remains uncertain and needed to be corrected.  The zero point was established by  shifting the average {\fuse}\ velocity of \ari\ and \oi\ MW absorption to the observed average heliocentric velocity of the MW component in the \hi\ 21-cm spectra. As \feii\ may be present in both the ionized and neutral gas, we compare the velocities of \feii\ at $\lambda < 1080$ \AA\ and $>1080$ \AA\ in order to ensure that the various segments of the detectors were aligned properly. The relative wavelength calibration for \fuse\ is often better than $\pm 5$ \km, but there is still sometimes some stretches on small wavelength scales that can be as large as $\pm 10$ \km. 

The STIS spectra have an excellent wavelength calibration, with a velocity uncertainty of $\sim$1--3 \km. The heliocentric velocity was then corrected to the dynamical local standard of rest-frame (LSR). We note that the MW \oi\ $\lambda$1302 and \hi\ 21-cm components align well, further ensuring that the MW \hi\ component can be used to align the \fuse\ spectra.

\subsubsection{Contamination and Continuum Placement}
As we are interested in lines that are shifted by about 100--180 \km\ in the LSR frame from their rest position, contamination from other lines may be problematic. However, a positive aspect of the velocity shift for the \oi\ lines is that the strong terrestrial airglow lines that may affect them, in particular \oi\ $\lambda$1039, are not an issue here.  For \oi\ $\lambda$1302, there is no contamination by any other features. The \oi\ transitions at $\lambda < 1000$ \AA\ were hopelessly contaminated in about 55\% of our sightlines (complicated stellar continuum, H$_2$ or \hi\ contamination). For the remaining 45\%, we were able to confidently use  \oi\ $\lambda$948 and/or $\lambda$936 in the LSR velocity interval $[+90,+180]$ \km. The other weaker transitions were rarely detected or often could not be used owing to a too complex stellar continuum.  

The HVC component of the \oi\ $\lambda$1039 line can be partially contaminated by the LMC component of the 5--0 R(2) H$_2$ $\lambda$1038.689 line. To estimate the contribution to the \oi\ HVC profile from this line, we fitted a Gaussian profile to the 4--0R(2) H$_2$ $\lambda$1051.498 line. This transition was picked because no other line contaminates it, the stellar continuum near this line can be easily modeled, and the strengths ($f \lambda$) of both transitions are about the same. Such a technique is further discussed and displayed in, e.g., \citet{lehner04}. When the blended H$_2$ line was not too strong, \oi\ $\lambda$1039 could be decontaminated and the resulting column density was in agreement within 1$\sigma$ uncertainty with that measured in the (uncontaminated) \oi\ $\lambda$$\lambda$948, 936 lines, giving us further confidence in our method to deblend the \oi\ $\lambda$1039 line.

Although \ion{Fe}{2} has many transitions in the {\fuse}\ wavelength range, the most useful was often the strongest transition available in the \fuse\ bandpass at 1144.938 \AA\ as it is uncontaminated by other interstellar lines and strong enough to be detected. However, in many cases we were also to be able to use weaker transitions at lower wavelengths. For the stars observed with STIS, we also made use of the \feii\ $\lambda$1608 line.

In order to make our measurements, we first needed to model  the stellar continuum near the absorption lines. The stellar continuum was often simple enough to be fitted with low-order Legendre polynomials ($\la 4$). However, in more complicated cases we fitted the stellar continuum with higher-order polynomials. In these cases, several continua were tested to be certain that the continuum error (see below) was robust \citep[see][]{sembach92}. In several cases we have estimates from multiple \oi\ and \feii\ lines that give similar column densities within 1$\sigma$ uncertainty, implying that the continua are reliable. For the cases where we do not have information from other lines, we are confident in our modeling because similar types of stars are used, which have similar overall continuum behavior. 

\subsubsection{Column Densities and Velocities}

To estimate the column density and velocities we adopted the apparent optical depth (AOD) method described in details by \citet{savage91}. In this method the absorption profiles are converted into apparent column densities per unit velocity $N_a(v) = 3.768\times 10^{14} \ln[F_c/F_{\rm obs}(v)]/(f\lambda)$ cm$^{-2}$\,(\km)$^{-1}$, where $F_c$ is the continuum flux, $F_{\rm obs}(v)$ is the observed flux as a function of velocity, $f$ is the oscillator strength of the absorption and $\lambda$ is in \AA\ \citep[atomic parameters were adopted from][]{morton03}.  The total column density was obtained by integrating over the absorption profile  $N = \int N_a(v)dv$. The values of the velocity, $v$, is obtained from $v = \int v N_a(v)dv / N_a $ \citep[see][]{sembach92}. The errors for the individual transitions include both statistical and continuum errors \citep[see][]{sembach92}. When no detection was observed, we estimated a 3$\sigma$ upper limit following the method described by \citet{lehner08} where the 1$\sigma$ equivalent width was estimated from integrating the continuum over a velocity interval $\Delta v$. If \oi\ or \feii\ was detected, we adopted $\Delta v$ from one of the detected species (this is valid as the profiles when both species are detected span a similar velocity extent). Otherwise we set $\Delta v = 50$ \km, a rough average based on sightlines where high-velocity absorption is detected. 

\citet{savage91} showed that the AOD method is adequate for data with $b_{\rm line} \ga 0.25$--$0.50 b_{\rm inst}$, where $b_{\rm line}$ is the intrinsic $b$-value of the line and $b_{\rm instr}$ is the $b$-value of the instrument. Since $b \equiv$\,FWHM$/1.667$, for {\em FUSE}, $b_{\rm inst} \approx 10$ \km. Therefore, we assume that a negligible fraction of the gas has $b\ll 2$ \km, or $T\ll 3800$ K if thermal motions dominate the broadening. Our assumption is supported by the fact that the \hi\ data show no narrow emission. As the peak apparent optical depth is generally less 1, we expect that saturation is not important.  Unresolved saturation can in fact be checked by comparing the apparent column densities estimated from the various transitions that have at least $\Delta [\log(f\lambda)] \ga 0.2$. Within 1$\sigma$, the apparent column densities of \oi\ $\lambda$$\lambda$1302, 1039, 948, 936, on one hand, and \feii\ $\lambda$1608, 1144, 1143, 1125, 1121, 1096, on the other hand, generally agreed. The exceptions were for sightlines where the high-velocity absorption of \oi\ $\lambda$1302 was such that $\tau_a\gg 1$ in the line center of the HVC component. Our final results  for the HVCs with $+90 \le v_{\rm LSR} \le +175$ \km\ are summarized in Table~\ref{t-sum}. The adopted column densities are a weighted average of the various estimates when more than one transition was available. 

\subsection{\hi\ Data and Measurements}
Two datasets were used for the \hi\ 21-cm data. First, we used the LMC \hi\ survey campaign undertaken by \citet[][hereafter S03]{staveley-smith03} using the Parkes 21-cm multibeam receiver \citep{staveley-smith96}. A full description of the data used in this work is available in S03. In short, the area covered was $13 \times 14$ deg$^2$ in RA and DEC, respectively, and centered on RA\,$=05^{\rm h}20^{\rm m}$ and DEC\,$=-68\degr 44\arcmin$ (J2000). The effective beamwidth was 14.3 arcmin. The original spectra were shifted to a common heliocentric reference frame, and were subsequently shifted to the Local Standard of Rest (LSR) frame. We used the data with the velocity spacing of the multibeam data of 0.82 \km, providing a resolution of 1.6 \km. The useful velocities range from about $-66$ to $+430$ \km, i.e.  the MW, HVC, and LMC components are covered.  Baselines were first adaptively fitted using polynomials of degree eight. Residual baselines were removed by fitting low-order polynomials. 

In Fig.~\ref{f-aod}, we show some ``typical" examples of \hi\ profiles compared to the \oi\ profiles. As it can be seen in this figure, the emission of the HVC is weak, and hence several baseline models were chosen, and the uncertainty in the baseline dominates the errors on the \hi\ column density and limits our sensitivity. The \hi\ column density was calculated by integrating the brightness temperature profile ($T_{\rm B}$) over the velocity range where HVC emission is observed, $N($\hi$) = 1.823\times 10^{18} \int T_{\rm B}(v) dv$ cm$^{-2}$, assuming that all of the photons emitted by the HVC escapes.  The average LSR velocities of the \hi\ profiles were estimated through $v = \int v T_{\rm B}(v)dv / \int T_{\rm B}(v)dv  $. The nominal column density sensitivity for the Parkes \hi\ observations is $8\times10^{16}$ cm$^{-2}$ across 1.6 \km\ based on the rms noise of 27 mK in the line-free region of the cube, but this does not take into account the uncertainty in the baseline. 

As the baselines of the earlier campaign remain somewhat uncertain, we also systematically used the more recent survey undertaken by \citet{mcclure09}, the Parkes Galactic All-Sky Survey (GASS). The resolution of these data is 1 \km\ and the spatial resolution is  16\arcmin. We refer the reader to McClure-Griffith et al. for a complete description of the survey. The important aspect of the data calibration is that the baseline quality is often more reliable. So even though GASS is about twice less sensitive than S03, we used these data to set a 5$\sigma$ limit on  $N($\hi$)$. We used the linewidth from the \oi\ absorption to estimate the limit on $N($\hi$)$ for each individual HVC. If \oi\ was not detected, we used the linewidth of \feii, and if \feii\ was not detected either, we set the linewidth to 50 \km. The \hi\ velocities and column densities are also summarized in Table~\ref{t-sum}. The values and errors of $N($\hi$)$ in this table were estimated using the two \hi\ datasets discussed above. We impose the condition that the HVC \hi\ 21-cm emission is detected in both surveys to set a detection.

\begin{figure}[tbp]
\epsscale{1} 
\plotone{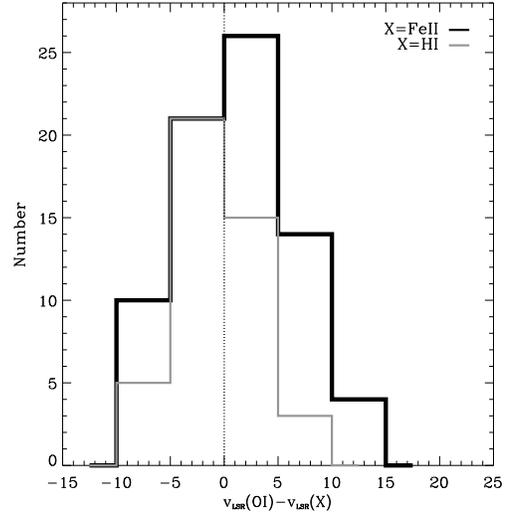}
\caption{Histograms showing the difference in the average LSR velocity of the HVC \oi\ profiles relative to \hi\ or \feii. Within the velocity calibration error and statistical uncertainties, $v_{\rm LSR}($\oi$)\approx v_{\rm LSR}($\hi$) \sim v_{\rm LSR}($\feii$)$. 
 \label{f-velcomp}}
\end{figure}

\begin{figure*}[tbp]
\epsscale{1} 
\plottwo{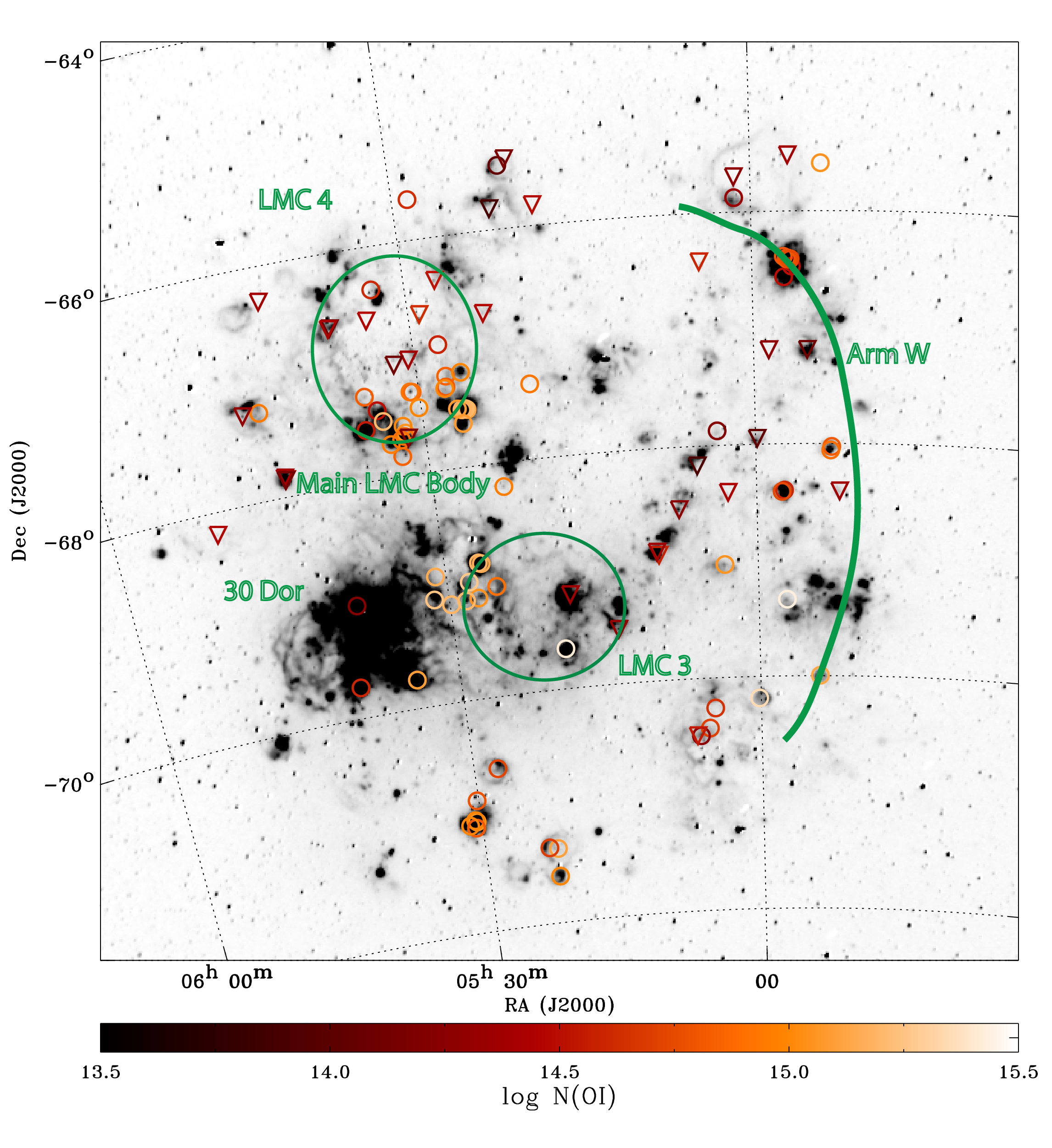}{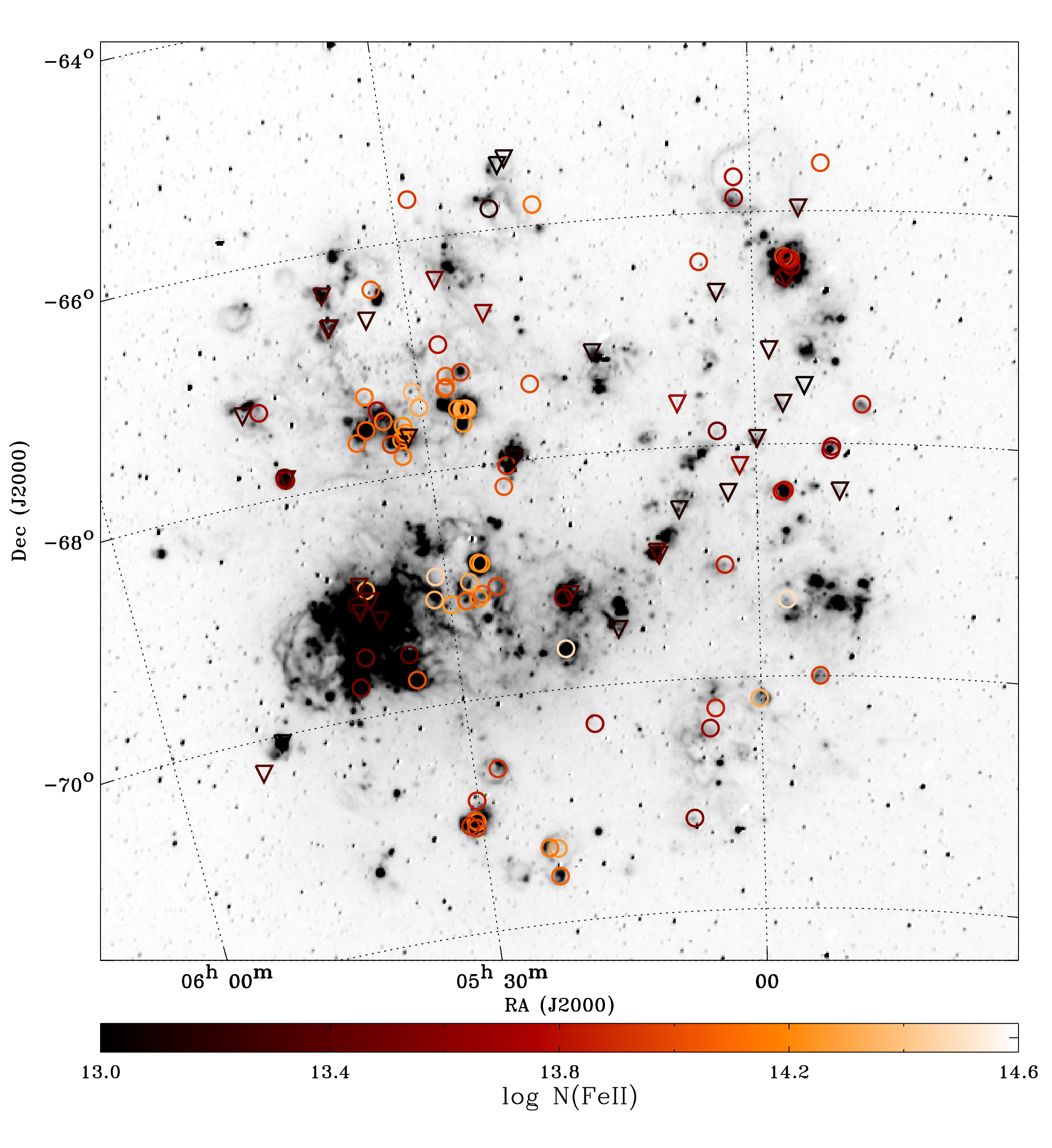}
\caption{H$\alpha$ image of the LMC \citep{gaustad01} where the darker regions correspond to brighter H$\alpha$ emission.  {\em Left:}\ The column density of \oi\ is overplotted where the color-scaling indicates the strength of \oi. The circles are detection of \oi\ and the  upside down triangles are 3$\sigma$ non-detection of \oi\ (and non-detection of \hi).  Only clouds with $90 \le v_{\rm LSR} \le 175$ \km\ are shown. We also show some rough schematics of prominent LMC features in green: the W arm and main body of the LMC \citep[see][]{staveley-smith03}; the supergiant shells LMC\,3 and LMC\,4 \citep[see][]{kim99}, and the unmistakeable 30 Dor region.  {\em Right:}\ Similar but for \feii. 
 \label{f-noi}}
\end{figure*}

\subsection{Estimates of $[$\oi/\hi$]$ and $[$\feii/\oi$]$}\label{sec-abund}

In the last column of Tables~\ref{t-sum}, we list the ratio \oi/\hi\ and \feii/\oi\ corrected for the solar oxygen abundance, where we use the standard square-bracket notation $[{\rm X^i/Y^j}] \equiv \log N({\rm X}^i) - \log N({\rm Y}^j) - \log ({\rm X/Y})_\odot $.  In this work we adopted the solar oxygen abundance $\log ({\rm O/H})_\odot = -3.27$ and iron abundance $\log ({\rm Fe/H})_\odot = -4.54$ recently recommended by \citet{lodders09}. We emphasize that the various profiles were integrated over a similar velocity range in order to compare the various species. Fig.~\ref{f-velcomp} shows that, within the errors, the average velocities of the \oi\ and \hi\ HVC profiles are aligned. For \feii\ and \oi, there is also a good agreement in most cases. There is evidence in several \oi\ and \feii\ profiles of at least two components separated by 20--40 \km\ (see Fig.~\ref{f-aod} and Figs.~5 and 6 in Lehner \& Howk 2007). In most cases the velocity structures of the  \oi\ and \feii\ absorption profiles follow each other, but there are cases where they do not (even though the velocity intervals of the profiles are the same). 
These differences suggest changes in the ionization structure with velocity along a given sightline, which can cause discrepancies between \oi\ and \feii\ average velocities. So in all cases, we compared the total \hi\ and \oi\ column densities, and apply the same rule when comparing \feii\ and \oi. 

The errors on $[$\feii/\oi$]$ include both the statistical and continuum placement, while the errors on $[$\oi/\hi$]$ include those and in addition the baseline uncertainties from the \hi\ data. The errors do not include the uncertainties on solar abundances  (see \S\ref{sec-disc} for more details on the solar O abundance).  In addition for $[$\oi/\hi$]$, there is a ``beam" error that arises from the comparison of data obtained from a 14\arcmin--16\arcmin\ beam compared to the pencil beam of the FUV observations. The $N($\hi$)$  represents an ``average" column density over a region near the line of sight toward the star. \citet{wakker01} studied this effect, and found that, for the HVCs, $N($\hi$)$ measured with a half-degree radio beam can differ by up to a factor 2--3 (either way) from the value measured with a 10\arcmin\ or 1\arcmin\ beam. The distribution of $N($\hi;36\arcmin$)/N($\hi;9\arcmin$)$ has a dispersion of about a factor 1.5, or 0.17 dex. Comparing $N($\hi$)$ measured with a 9\arcmin\ beam or with a 1\arcmin\ beam or through Ly\,$\alpha$ absorption gives 0.06 dex. Using Wakker et al.'s results, \citet{lehner04} estimated  a beam error of 0.10 dex for a 21\arcmin\ beam and 0.06 dex for a 12\arcmin\ beam. We therefore estimate that Parkes 14\arcmin--16\arcmin\ data have a beam error of about $\pm 0.08$ dex.

As there are fields of view with several stars within 14\arcmin--16\arcmin, we can test the validity of the estimated beam error using the \oi\ column densities. For example, near the field defined by RA\,$ =  (4.945 \pm 0.005)^{\rm h}$ and DEC\,$=(-66.415 \pm 0.005)\degr$ intervals, there are 9 estimates of  $N($\oi$)$ but only a single \hi\ pointing. The mean column density of \oi\ and dispersion are  $10^{14.80 \pm 0.06}$ cm$^{-2}$. The dispersion is consistent with a beam  error of $\pm 0.08$ dex.  In Fig.~\ref{f-noi}, we show the spatial distribution of the \oi\ column density projected on an H$\alpha$ map of the LMC. While on large-scale structure there is variation in $N($\oi$)$, on small scales ($<15\arcmin$), there is generally no evidence of variation other than noise fluctuation. An exception is toward BI272 and Sk--67\D266 that are separated by 8.6\arcmin\ (and the \hi\ pointings are apart by 9.1\arcmin) where the \oi\ columns are different by at least 0.6 dex, but in that case the \hi\ column densities are also likely very different (along Sk--67\D266 no \hi\ is detected but it is toward BI272). This may be an example where the edge of a cloud is probed. In summary, statistically, a beam error of 0.08 dex appears to be adequate for the present sample.

\section{Properties of the HVCs toward the LMC}\label{sec-prop}

\subsection{Sky Covering Factor and Multiphase Gas}\label{ssec-sky}

The detection rate of high-velocity ($90 \le v_{\rm LSR}\le 175$ \km) \oi\ absorption (based on \oi\ $\lambda$1039) is 70\% (78/111), while for \feii\ (based on the transition at $\lambda$1144) it is slightly higher, 72\% (93/129). In contrast for the present sample, the detection rate of the HVC \hi\ 21-cm emission is only 32\% (44/139). The difference between emission and absorption measurements is simply due to the much higher sensitivity of the absorption data that scale linearly with the density. While \oi\ probes solely the neutral gas, \feii\ can trace both the neutral and ionized gas. An examination of the \fuse\ profiles along each lines of sight where \oi\ and \feii\ are detected also shows high-velocity absorption in the profiles of \nii\ $\lambda$1083 and \ovi\ $\lambda$1031  \citep[for examples of \ovi\ profiles, see][]{lehner07,howk02}, which are tracers of photoionized and collisionally ionized gas, respectively. Although we do not provide a detailed analysis of these species in this work, a systematic inspection of their profiles shows that the detection rate of \nii\ is about 70\% as well, and for \ovi\, the incidence could be about 90\% \citep[a detailed study will be needed to confirm the number for \ovi\ as H$_2$ can contaminate the $\lambda$1031 transition,][but we note that the incidence of HVC \ovi\ in the Lehner \& Howk sample -- 19 stars -- is 100\%]{howk02,lehner07}, implying a substantial presence of weakly and highly ionized gas. We have also already noted that the velocity centroids and velocity extents of the \oi\ and \feii\ profiles generally overlap (see Fig.~\ref{f-velcomp}), and this also applies for \nii\ and \ovi. The STIS spectra also show absorption in \siiii, \civ, and \siiv\ at the HVC velocities \citep[see][]{lehner07}. This implies that the HVC is multiphase, with several temperature and ionization regimes coexisting kinematically.  However, we did not find any evidence for H$_2$ absorption. To do our search we systematically constructed H$_2$ velocity profiles for key transitions. We looked at the same H$_2$ transitions as those used in low H$_2$ column density sightlines \citep[see, e.g.,][]{lehner02,richter01}.  In view of the largely ionized gas and low $N($\hi$)$ (see \S\ref{ssec-oh}), the absence of molecular gas is not surprising. 

In Fig.~\ref{f-noi} we show the column density of \oi\ (left-hand side) and \feii\ (right-hand side) of the HVCs for each sightlines overplotted on an H$\alpha$ map \citep[from][]{gaustad01} of the LMC. This map shows that the HVCs are projected on both active (superbubbles, supergiant shells) and quieter (field stars, diffuse \hii\ region)  regions. The stars are also situated in various regions of the LMC, such as the spiral arm W and its main body. \oi\ and \feii\ absorption at  $+90\le v_{\rm LSR} \le +180$ \km\ are detected toward both active and quiet regions as well as spiral and central regions. As already noted in  \S\ref{sec-abund}, there is no strong evidence for a large variation in $N($\oi$)$ or $N($\feii$)$ on small scales. The variation occurs on several tens of arcminutes or  degrees. The variation does not appear correlated with the projected physical regions as both active and quiet ones can have both high and low $N($\oi$)$ or  $N($\feii$)$. Hence the overall variation appears more related to the patchiness of the HVCs and/or projection effects and/or changes in the ionization structures (e.g., the HVC gas may be more highly ionized gas in some directions than others). Nevertheless the high detection rate of high-velocity absorption implies that the HVCs cover a substantial area of the LMC.

\subsection{Distribution of the Average Velocities}

\begin{figure}[tbp]
\epsscale{1} 
\plotone{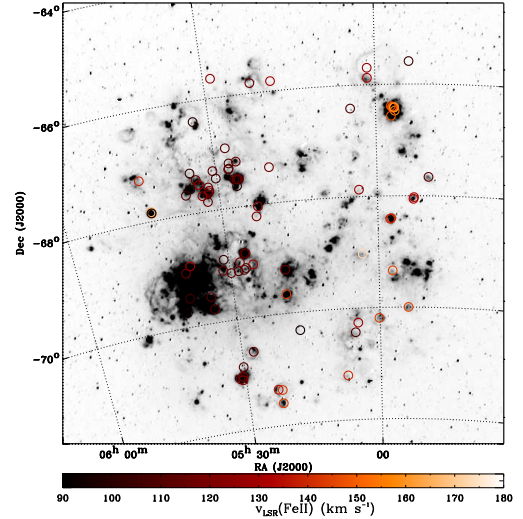}
\caption{Similar to Fig.~\ref{f-noi}, but here the HVC LSR velocity of \feii\ is overplotted onto the H$\alpha$ image of the LMC. The color-scaling indicates the velocity of the absorption feature.  Note the absence of trend with the declination, but a grouping in velocities with the  right ascension: At RA\,$< 5.2^{\rm h}$, the average LSR velocity  is $140 \pm 17$ \km, while at RA\,$> 5.2^{\rm h}$, it is $114 \pm 14$ \km.  \label{f-vfe}}
\end{figure}

In Fig.~\ref{f-vfe}, we show the average LSR velocity the \feii\ HVC absorption for each sightlines overplotted on the H$\alpha$ map of the LMC.  In contrast to the column densities, a few trends can be noted in the distribution of the velocities with coordinates. In most regions, there is little variation in the velocities on small scales. The second trend is that the HVC can be separated in roughly two  LSR velocity zones: At RA\,$< 5.2^{\rm h}$, the average LSR velocity  is $140 \pm 17$ \km, while at RA\,$> 5.2^{\rm h}$, it is $114 \pm 14$ \km. Most of the HVCs at RA\,$< 5.2^{\rm h}$ are roughly projected on the ``W" arm of the LMC, while at larger RA, they are projected on the main body of the LMC and regions with many supergiant shells. 

\begin{figure}[tbp]
\epsscale{1} 
\plotone{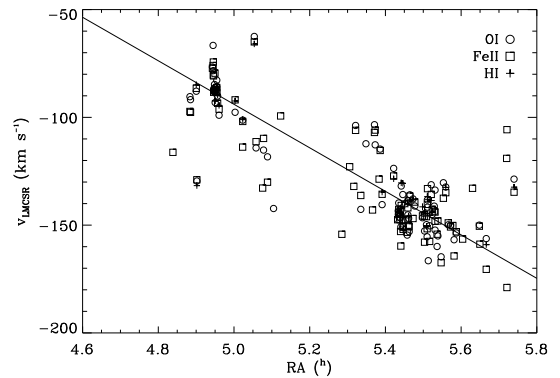}
\caption{Comparison of the LMC standard of rest  average velocities of the HVCs derived from \oi, \feii, and \hi\ with the right ascensions. The solid line is a linear fit to the data.  \label{f-sumv}}
\end{figure}

As the (Galactic) LSR frame may introduce some spurious velocity gradients, we also estimated the velocities in the LMC standard-of-rest (LMCSR) frame using the velocity vector of the LMC, ${\bf v}_{\rm LMC} =( -86 \pm -12, -268\pm 11, 252 \pm 16) $ \km\ \citep{kallivayalil06}:
$$
v_{\rm LMCSR} = v_{\rm GSR}+86 \cos l \cos b + 268 \sin l \cos b - 252 \sin b,
$$
where  the Galactic standard-of-rest frame is $v_{\rm GSR} = v_{\rm LSR} + 220 \sin l \cos b$, and 220 \km\ corresponds to the velocity of the solar circular rotation around the MW.\footnote{There might be a 15\% increase in the MW circular velocity, i.e. 251 \km\ instead of 220 \km \citep{shattow09}. Combining the high circular velocity with the proper motions derived by \citep{piatek08} (implying ${\bf v}_{\rm LMC} =( -83, -243, 238) $ \km) gives essentially the same distribution.}  Fig.~\ref{f-sumv} shows an anti-correlation between the LMCSR velocity and the RA. Although there is some scatter in this figure, there is little doubt of a general increase in the $| v_{\rm LMCSR}|$ with increasing RA (or with the galactic latitude), which is highlighted by the solid line, a linear fit to the data using a ``robust" least absolute deviation method.  No pattern is observed with the declination (see Fig.~\ref{f-vfe}). 

This apparent velocity gradient with the RA is unlikely to be caused by the underlying motion of the gas in the LMC disk.  Indeed, on average for the sightlines included in our sample, the difference in the LMC disk average velocities (estimated using the \hi\ emission data) between the two RA zones is small. Hence this gradient may be more related to phenomena producing these HVCs rather than the disk-gas motions of the LMC (assuming these HVCs have their origin in the LMC, a hypothesis that appears valid as we discuss in \S\ref{sec-disc}). And indeed, the RA\,$> 5.2^{\rm h}$ zone is where 30 Dor and most of the supergiant shells are \citep[see Figure~2 in][]{kim99}, regions likely to  produce stronger galactic outflows than in the spiral arm.  In \S\ref{sec-disc} we further discuss in more details the implications of this difference in the kinematics.

\subsection{Absolute and Relative Abundances}\label{ssec-oh}
\begin{figure}[tbp]
\epsscale{1} 
\plotone{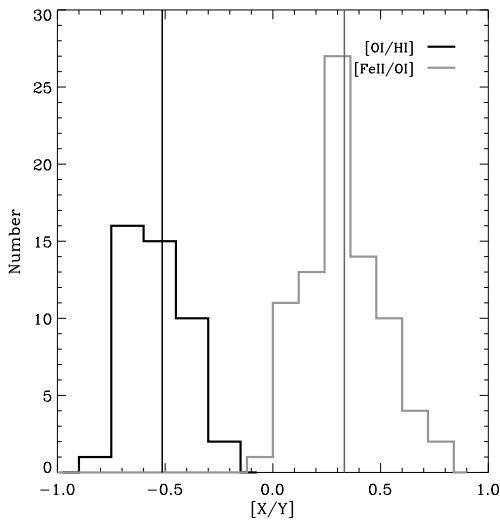}
\caption{Distribution of $[$\oi/\hi$]$ and $[$\feii/\oi$]$ of the HVCs. Limits are not included.  The vertical solid lines show the mean values for $[$\oi/\hi$]$ and $[$\feii/\oi$]$. 
 \label{f-sumplot}}
\end{figure}

In Fig.~\ref{f-sumplot}, we show the distributions of $[$\oi/\hi$]$ and $[$\feii/\oi$]$. The histograms include all the measurements but not the limits. The median, mean, and 1$\sigma$ dispersion for $[$\oi/\hi$]$ are $-0.53$ and $-0.51 \,^{+0.12}_{-0.16}$ (44 data points),\footnote{All the mean and dispersion values of the abundances are calculated in the linear scale and then converted to the logarithmic scale.} respectively, and for  $[$\feii/\oi$]$ are $+0.30$ and $+0.33 \,^{+0.14}_{-0.21}$ (77 data points). With the Kaplan-Meier product limit estimator method that can be used to determine the mean of a set of data points containing limits \citep[e.g.,][]{feigelson85,isobe86}, we find $[$\oi/\hi$] = -0.45$, entirely consistent with the result above. For  $[$\oi/\hi$]$, the dispersion is similar to the statistical plus systematic uncertainties, and hence the scatter is likely mostly due to the measurements. However, for $[$\feii/\oi$]$, the errors in the individual sightlines are much smaller than the dispersion, implying that those are real variations due to changing conditions in the ionization and/or dust depletion (see below). 

In Fig.~\ref{f-sumplot1}, $[$\oi/\hi$]$ and $[$\feii/\oi$]$ are compared to $N($\oi$)$. For $[$\oi/\hi$]$, we also show the lower limits (i.e. when \hi\ is not detected). There is a lack of data at  $14.8 \la \log N($\oi$) \la 15.3$  and  $[$\oi/\hi$]<-0.8$ or  $[$\oi/\hi$]>-0.2$, indicating that neither very subsolar nor solar absolute abundances are found in these HVCs. We emphasize that if HVC gas with $[$\oi/\hi$]<-0.8$ were common, it should have been easily detected as $N($\hi$) > 10^{18.9}$ cm$^{-2}$ whereas gas with  $[$\oi/\hi$]>-0.1$ would imply $N($\hi$) < 10^{18.5}$, typically smaller than detected. As we discuss in \S\ref{sec-disc}, the abundance of oxygen is consistent with the O abundance in the LMC, suggesting that the origin of these HVCs is gas that has been ejected from the LMC. We did not find any discernable patterns between  $[$\oi/\hi$]$ and RA or DEC or velocities.

\begin{figure}[tbp]
\epsscale{1} 
\plotone{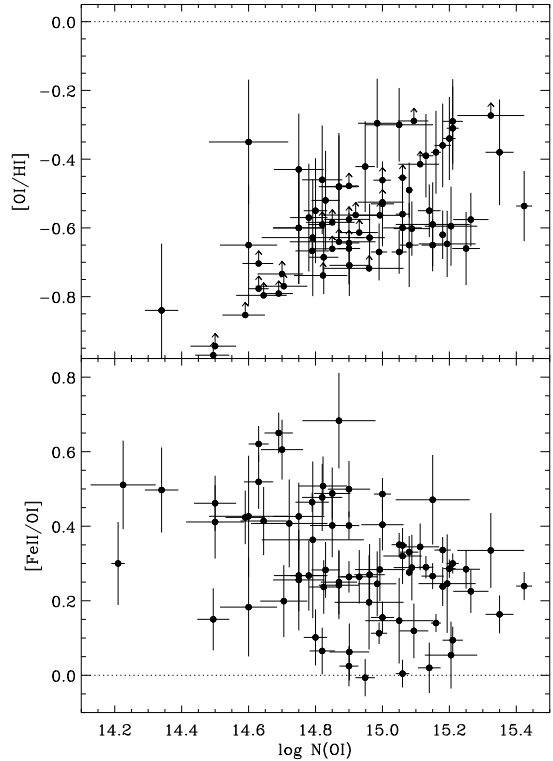}
\caption{$[$\oi/\hi$]$ and $[$\feii/\oi$]$ of the HVCs against the HVC \oi\ column density. 
 \label{f-sumplot1}}
\end{figure}

In order to correctly interpret the $[$\feii/\oi$]$ ratio, we need to know if chemical evolution or dust depletion (besides ionization) plays a role when comparing Fe and O  \citep[e.g.][]{savage96,jenkins04,sofia04,howk06}. In the MW gas, refractory elements like Si and Fe are more depleted from the gas phase than S or P (or O) if dust grains are present. In Galactic halo-like clouds, we would expect that $[{\rm Fe/O}] \simeq -0.4$ if ionization was unimportant. Chemical evolution of the gas can also affect the relative abundances. For example, while the bulk of O is produced in massive stars ($>12$ M$_\odot$), only about a 1/3 of Fe is produced in the remnants of these massive stars; the bulk of Fe is thought to be made in Type Ia supernovae \citep[e.g.,][]{russell92}. Hence, if some gas was recently enriched by a supernova, $[{\rm Fe/O}]$ would be expected to be subsolar. And indeed,  supernova SN 1987A produced a ratio $[{\rm Fe/O}] \sim -0.6$ dex solar \citep{dopita90}, unfortunately a level quite similar to dust depletion, implying a degeneracy in the possible origins of the $[$Fe/O$]$ deficiency. The $\alpha$-elements (e.g., S, Si), on the other hand, are believed to follow the chemical evolution of O. So if Si is depleted relative to (undepleted) S, this should result mainly from a dust depletion effect (assuming the ionization effect is negligible).

Besides the \feii/\oi\ ratio, only toward a few sightlines can we do an analysis of the relative abundances of singly ionized species because in most cases the S/N is not high enough to detect species other than \feii\ and \oi\ (and saturated \nii) in the \fuse\ spectra. However, the 10 STIS sightlines allow us to access to stronger transitions, such as \feii\ $\lambda$1608 and \siii\ $\lambda$1526, as well as the undepleted element S, via  \sii\  $\lambda$$\lambda$1253, 1259. In Table~\ref{t-rel} we summarize the relative abundances $[$\siii/\oi$]$, $[$\siii/\sii$]$ (or $[$\siii/\pii$]$ for the 4 \fuse\ sightlines), and $[$\feii/\siii$]$ estimated in the HVCs along the 10 STIS sightlines plus the 4 \fuse\ stars with the highest S/N spectra. As Si and O follow the same chemical evolution, only ionization and depletion should affect $[$\siii/\oi$]$. If a mild dust-depletion exists in the HVC gas, $-0.2 \la [$\siii/\oi$] \la 0$ would be expected. As we observe $+0.2 \la [$\siii/\oi$] \la +0.7$ (with a mean of $[$\siii/\oi$] = +0.48 \,^{+0.15}_{-0.25}$), this implies that ionization is extremely important and its effect dominates over depletion. Si and S also follow the same chemical evolution, so again only depletion and ionization may affect this result. However, the ionization effect should be milder as two singly-ionized species are compared. In a few cases, there seems to be a mild depletion of Si at the level of 0.1--0.2 dex based on the comparison of Si and S. However, toward the X-ray binary 4U0532-664, $[$\siii/\sii$]> +0.23$ (it has also one of the highest $[$\siii/\oi$]$ ratios) suggests that ionization effects cannot be neglected even when only singly-ionized species are compared. Hence the intrinsic dust depletion of Si may be larger. Finally, $[$\feii/\siii$]$ varies between a solar value and $-0.3$ dex solar. This is somewhat consistent with a Galactic halo-like depletion pattern observed in the MW \citep[e.g.][]{savage96,howk99}. We note that according to models of diffuse ionized gas \citep{sembach00}, \feii, \pii, and \sii\ can be more affected by ionization than \siii. Unfortunalely neither \feiii\ nor \Siii\ can be estimated reliably and accurately. It is, however, beyond doubt that ionization is important as illustrated independently by the high detection rates of \nii\ and \ovi\ (see \S\ref{ssec-sky}). Following the conclusion from $[$\siii/\sii$]$, the mean value $[$\feii/\oi$] = +0.33 \,^{+0.14}_{-0.21}$ implies that the HVC gas is very ionized toward every directions probed in our sample, $> 50$\% on average but could be as large as $>90$\% toward many directions. Even if ionization effects dominate the relative abundances, there is evidence of dust depletion in these HVCs in view of the subsolar  $[$\siii/\sii$]$ ratio toward several sigthlines. In view of the Si depletion, it is extremely likely that Fe is depleted as well, and nucleosynthesis must play only a little role (if any) in the Fe defiency in these HVCs.

\section{Origin the HVCs toward the LMC}\label{sec-disc}

We now offer some comments on the implications of the kinematics, ionization, and abundances of the HVCs for their origin. The first implication is that the HVCs across the LMC are extremely likely to be an HVC complex, i.e. they have the same origin. We draw this conclusion from both kinematics, metallicity, and ionization arguments. In the previous section, we saw there is no strong evidence for a large variation in the metallicities of the HVCs over the face of the LMC (see Fig.~\ref{f-sumplot}).  While the HVCs cover a large interval of LSR velocities between $+90$ and $+175$ \km, there seems to be a gradient in the velocity with the RA across the LMC. This contrasts from earlier studies that use smaller samples and found no pattern \citep[see][]{savage81,danforth02}.  Along all sightlines, the HVCs are largely photoionized ($>50$--80\%) and also collisionally ionized (based on the high incidence of high-velocity \ovi\ absorption). These results strongly suggest that these HVCs are part of a general complex of dominantly ionized gas moving at high velocity relative to both the LMC and the MW. The \hi\ 21-cm contours shown in Fig.~8 of S03 further support this conclusion. The neutral part of this HVC complex covers the sky from DEC\,$\sim -74 \degr$ to $-65 \degr$ and RA\,$\sim 4^{\rm h}$ to  $6^{\rm h}$ based on the combined \oi\ and \hi\ observations, and hence the complex covers a large portion of the LMC area. In our sample, the \hi\ column densities range from $<10^{18.4}$ ($<10^{16.6}$ cm$^{-2}$ if \oi\ is used as a proxy for \hi\ after correcting for the metallicity) to about $10^{19.2}$ cm$^{-2}$, and therefore these HVCs are possible local analogs of the Lyman limit systems seen at higher redshifts in the spectra of QSOs.

Recent works have presented some indirect evidence that the HVCs seen toward the LMC are likely related to outflows from the LMC. \citet{staveley-smith03} found that the peak \hi\ column densities are seen projected onto \hi\ voids (such as supergiant shells, e.g., LMC\,3) within the LMC disk and furthermore are connected to the disk of the LMC with spatial and kinematics bridges in position-velocity plots, suggesting they are ejection of matter from the LMC disk. \citet{lehner07} found a systematic similarity in the \ovi/\civ\ ratio between the LMC and HVC components  (which were quite dissimilar from the IVC and MW components), suggesting again the HVC has its origins in the LMC. In the localized region of 30 Dor, \citet{redman03} presented high resolution emission of [\oiii] showing  high-velocity features at similar velocities that the HVCs studied here, which again can be traced back to the 30 Dor regions in position-velocity plots, and hence likely trace the superwind of 30 Dor. Using a Gaussian decomposition of the \hi\ 21-cm, \citet{nidever08} argued that supergiant shells in the southeast \hi\ overdensity region of the LMC are blowing out gas and energy. Below we argue that our results combined with these findings leave little doubt that the LMC has indeed strong outflows of matter and energy from stellar feedback occurring within its disk. 

The metallicity of the HVC and the distance to the HVC are the two most direct and important quantities for deciphering the origin of a given HVC \citep[e.g.,][]{wakker01,zech08}. The distance of this HVC complex is likely to remain unknown, leaving us only the metallicity information. In \S\ref{ssec-oh} we show that the mean oxygen abundance for the HVCs in our sample is  $[{\rm O/H}] =-0.51 \,^{+0.12}_{-0.16}$. By definition, $[{\rm O/H}]$ is relative to the solar O abundance, so  this abundance rules out the origin of this gas from outflows within the solar circle.  However, is it possible the gas could have  originated from the outer regions of the Galactic disk, since the abundances of the MW seem to vary as a function of the galactocentric  radius, $R_{\rm g}$. Several studies suggest indeed a slight gradient in ${\rm O/H}$ from $-0.01$ to about $-0.07$ dex\,kpc$^{-1}$ \citep[e.g.,][and references therein]{daflon04,esteban05,rudolph06}. As there is still much uncertainty in the gradient value itself (and the value could vary with $R_{\rm g}$), we follow \citet{cescutti07} who divided the data in various bins of $R_{\rm g}$ and find that the lowest $[{\rm O/H}]$ has a mean and a deviation of $-0.2\pm 0.2$ dex at $R{\rm g} > 9.5$ kpc (at smaller $R_{\rm g}$, $[{\rm O/H}]>-0.1$), which is still notably larger than the metallicity of the HVC complex (even when depletion is taken into account, see below). Hence if the HVC had its origin from a Galactic fountain phenomenon, its metallicity would require to have originated in the outer regions of the MW where a large cluster of massive stars with much subsolar metallicities than found elsewhere in the MW would need to be present. This seems unlikely and to the best of our knowledge there is no observational evidence of such a low-metallicity cluster of massive stars in the MW disk. Therefore the metallicity of HVC does not support a MW origin and combined with the other observational findings discussed above and below it seems safe to reject the MW hypothesis as the origin for the HVC complex. 

The next two more likely candidates for the origin of the HVC complex are the LMC and SMC. In order to compare the abundances, we first need to know the present-day O abundances of these galaxies. As there is currently no reliable estimate of the interstellar oxygen abundance in absorption in these galaxies, we use for comparison the stellar and nebular abundances. The recent results from the analysis of 133 and 80 B-type stars in the LMC and SMC yield a present-day mean oxygen abundance of $-3.66\pm 0.13$ and $-4.01\pm 0.21$ dex in the LMC and SMC, respectively \citep{hunter09}. These values are quite consistent with the O abundances estimated in \hii\ regions within the SMC and LMC \citep{kurt98}. The average of the values derived in the \hii\ regions and B-type stars yields $-3.64 \pm 0.10$ dex for the LMC and $-4.00 \pm 0.16$ dex for the SMC. In the last 20 years, the solar abundance of oxygen has changed by nearly 0.2 dex \citep[e.g.,][]{anders89,grevesse96,asplund06}. Here we adopted the most recent recommended O abundance of $-3.27$ dex \citep{lodders09} (an intermediate value between the low and high estimates), which is an average value from the recent estimates of \citet{caffau08}, \citet{ludwig08}, and \citet{melendez08}. As we use it as a systematic reference, the uncertainty in the solar O abundance does not affect our conclusions. The above LMC and SMC abundances relative to solar abundances are $-0.40 \pm 0.10$ and $-0.73 \pm 0.16$ dex solar, respectively. These abundances are consistent with those measured from the LMC and SMC Cepheids \citep{luck98} and there is no evidence for a large abundance variation in these galaxies. 

Although there is some overlap with the SMC abundance,  the mean and the dispersion of O/H of the HVC is closer to the canonical LMC value. Furthermore several lower limits are well above the SMC value and approach the LMC O abundance. This strongly favors an LMC  origin for the HVC complex. As we compare \oi\ with \hi, ionization does not affect this ratio. However, dust depletion of oxygen could affect HVC O/H abundance, albeit by no more than $0.1$ dex. Indeed, in the Galactic disk the evidence for oxygen dust depletion is found to be $\la 0.1$ dex with the adopted solar oxygen  abundance \citep{meyer98,andre03,cartledge04,jensen05}. It is currently unclear if the O depletion is different in the more diffuse gas than in the Galactic disk and it is often assumed to be the same \citep[e.g.,][]{welty99} and in any case it is likely less or equal to the Galactic disk value based on the depletion patterns from other elements. In \S\ref{ssec-oh}, we argue that there is some evidence of dust depletion in these HVCs. Hence if oxygen is depleted by  about $0.05$--$0.1$ dex, the average HVC O abundances listed above would converge to the canonical LMC O abundance.  A scenario where a pre-existing low metallicity gas may have reduced the overall abundance of the HVC seems less compelling because this would require that the HVC had mixed with the (unknown) low-metallicity component and had time to fully mix as there is no evidence of extremely low-metallicity gas in these HVCs. 

Therefore the HVC metallicity is consistent with an origin mostly from the LMC itself in form of outflowing matter.   An alternative scenario would be that the HVC is the result of a hydrodynamical interaction between the LMC and the hot halo of the MW, which, via ram pressure, could remove gas from the LMC disk. In this case, the HVC would also be expected to have an abundance similar to the present-day LMC. However, such models rule out any LSR velocities at 100--180 \km\ for the ram-pressure gas near the center of the LMC; only $40\degr$ away from the LMC center, ram-pressure models produce gas at LSR velocities  100--160 \km\ \citep[see Fig. 12 in][]{mastropietro05}. 

The multiphase structure of the HVC complex, the large fraction of ionized gas, and the presence of dust grains are all also consistent with an energetic galactic outflow. Galactic winds are indeed known to be extremely multiphase and to entrain dust grains far away from the galactic disk \citep[e.g.][]{veilleux05,strickland07,howk09}. The presence of a diffuse hot gas and shock heated and ionized gas would be expected in galactic outflows \citep[e.g.,][]{veilleux05}. The high ions (\ovi, \civ, \siiv) are signatures of such collisionally ionized gas and  suggest the presence of a diffuse hot ($>10^6$ K) phase interacting with the cooler ($10^4$ K) neutral and ionized gas \citep[see discussion in][]{lehner07}. This HVC complex also shares similar ionization properties to those of the complex multiphase structures of the HVCs toward the globular cluster M\,5 in our MW, HVCs believed also to probe energetic events that occur in the central region of the MW \citep{zech08}. 

Finally, the observed velocity gradient of the HVC across the LMC can also be simply explained in the context of outflows from the LMC.  Figure~2 in \citet{kim99} shows a larger concentration of supergiant shells at RA\,$>5.2^{\rm h}$ than at lower RA. The combined effect of these supergiant shells must produce stronger outflows in the central region of the LMC than in one of its arm. The high velocities of the gas observed toward 30 Dor \citep{redman03} are also quite consistent with the higher velocities observed at RA\,$>5.2^{\rm h}$, strongly suggesting that 30 Dor (unsurprisingly) participates to the overall outflow from the LMC. 

In order to know if those outflows can escape the gravitational potential of the LMC, the rotation velocity of the LMC needs to be known. Unfortunately, the rotation curves using different methods and tracers allow a large interval of rotational velocities, from about 70 to 107 \km\ \citep[e.g.,][]{alves00,olsen07,vandermarel09}, which would imply escape velocities from the LMC gravitational potential from $\sim$100 to $\sim$150 \km. Results based on the LMC proper motions yield an even higher $v_{\rm rot} \simeq 120$ \km\ \citep{piatek08}, and hence higher escape velocity. In many cases the absolute LMCSR velocities of the HVC range from 100 to 170 \km\ (see Fig.~\ref{f-sumv}). The tangential velocities and inclination of the outflows are not known, but in view of such large radial velocities, it seems reasonable to conclude that the LMC may have a galactic wind as long as $v_{\rm rot} \la 100$ \km. Furthermore, if the HVC gas does not escape the LMC potential, there should be signatures of infalling material. We searched for high positive velocity absorption relative to the LMC disk and only found less than a handful sightlines with \feii\ absorption at $+50$--70 \km\ relative to the LMC disk, possible signatures of infalling gas. The two most compelling sightlines are shown and discussed in \citet{howk02}. Therefore as these authors concluded with their smaller sample, there is no strong evidence of infalling material in these directions, and hence the HVC complex does not seem to be the result of an outflow associated with a galactic fountain and hence is more likely a galactic wind.  At 150 \km, ignoring possible drag forces and initial higher velocities, it would take about 30 to 65 Myr to reach a distance of 5 to 10 kpc, respectively. Therefore a small galaxy (1--$3\times 10^{10}$ M$_\odot$) such as the LMC is capable of producing outflows that may pollute its surroundings beyond its gravitational potential. 

The mass of the neutral gas in the outflow can be roughly estimated  via,
$$
M^{\rm neut}_{\rm HVC} = f_{\rm cov} \frac{\overline{N}({\rm O\,I})}{({\rm O/H})_{\rm HVC}} \mu_{\rm H} m_{\rm H} A_{\rm HVC}, 
$$
where the covering factor is $f_{\rm cov} = 0.7$ from our detection rate, the average \oi\ column density is $\overline{N}({\rm O\,I}) \simeq 10^{14.9}$ cm$^{-2}$, $m_{\rm H} = 1.673 \times 10^{-24}$ g  is the atomic mass of hydrogen, and $\mu_{\rm H} \simeq 1.3$ corrects for the presence of helium. The HVC seen in absorption covers the LMC disk, so we take the area of the HVC, $A_{\rm HVC}$,  to be similar to  the disk of the LMC assuming a diameter of 7.3 kpc, the diameter of the \hi\ disk \citep{kim98}. This yields $M^{\rm neut}_{\rm HVC} \sim 1.5\times 10^6 $ M$_\odot$. Using the \hi\ data with $\log N($\hi$)>17.4$, we estimate an \hi\ mass of $0.4 \times 10^6$ M$_\odot$ for the HVC gas, which gives after correcting for helium $0.5 \times 10^6$ M$_\odot$. The main difference between these values are the difference in the covering factors between absorption and emission data. The absorption method also likely overestimates the mass because it assumes the same \hi\ column density while the \hi\ emission estimate misses the lower column density gas, and therefore we just adopt the average value of $\sim 10^6$ M$_\odot$. These estimates also assume that the HVC complex is at 50 kpc. If  the HVC is 10 kpc away from the LMC toward us, and assuming no projection correction, the mass would be $0.6 \times 10^6$ M$_\odot$. The neutral mass is, however, only the tip of the iceberg as the HVC is largely photoionized and collisionally ionized. Including the photoionized high-velocity gas probed by \feii\ and \nii, the mass of the HVC increases by at least a factor 2--3 (see above). Using the measurements of the \ovi\ absorption in the HVC component \citep{lehner07}, the average \ovi\ column density (based on 4 measurements) is $10^{13.9}$ cm$^{-2}$, which would yield a \hii\ column density for the collisionally ionized gas of $N_{\rm coll}($\hii$) \approx 10^{19.3}$ cm$^{-2}$ assuming that the metallicity is the same as the neutral HVC and the \ovi\ ion fraction is 0.2. The latter value cannot be much larger than 0.2, but could be much smaller if the temperature of the gas is far (either way) from the peak temperature of \ovi\ in collisionally ionized equilibrium \citep{gnat07}, hence  $N_{\rm coll}($\hii$) >10^{19.3}$ cm$^{-2}$. Therefore the total mass of the outflow is more likely to be  $>(0.5$--$1)\times 10^7$ M$_\odot$. While this mass is much smaller than the total masses of the Stream and Bridge \citep{bruns05,lehner08}, it is similar to or possibly even larger than other large HVCs such as Complex C \citep{wakker01a,wakker07,thom08}. In view of our results, it is plausible that Complex C might be the remnant of an outflow from a hidden dwarf galaxy that is being accreted by the MW. The same fate could await the LMC HVC complex.

\section{Summary}\label{sec-conc}

Using the \fuse\ LMC database, we search for absorption at  high LSR velocity ($+90 \le v_{\rm LSR} \le +175$ \km) of  neutral, ionized, and molecular species in the spectra of  early-type stars located in the LMC. We use sensitive Parkes observations of \hi\ 21 cm for comparison, and for 10 stars, we completement the \fuse\ data with STIS E140M data.   Our final sample consists of 139 lines of sight, 44 of them showing both \oi\ absorption and \hi\ emission at high LSR velocities, an unprecedented number of sightlines where the metallicity can be estimated in a large HVC complex. Our main results for the HVCs toward the LMC are as follows:

1. The detection rates of high-velocity \oi\ $\lambda$1039,  \feii\ $\lambda$1144, and \nii\ $\lambda$1083 absorption toward our LMC stars are about 70\%. The detection rate of \ovi\ $\lambda$1031 could be even higher, $\sim$90\%.  While the quality of the data changes from sightline to sightline, the high detection rates imply that the HVCs cover a subtantial area of the LMC. In contrast the \hi\ 21-cm emission spectra along the same sightline only reveal the HVC at a 32\% rate.  

2. Since the neutral, ionized, and highly ionized HVCs are found at similar velocities along a given sightline, the HVC has a complex multiphase structure. However, we did not find any evidence of cold ($T<10^3$ K) or molecular gas, which is not surprising in view of the low \hi\  column density of the HVCs ($\log N($\hi$)\la 19.2$). 

3. We find a mean and dispersion of $[$\siii/\oi$] = +0.48 \,^{+0.15}_{-0.25}$ (14 sightlines). Since Si and O have similar chemical evolution and are only mildly depleted in diffuse gas, this implies that the gas is  $>50$--80\% ionized. Using a larger sample, we find  $[$\feii/\oi$] = +0.33 \,^{+0.14}_{-0.21}$ (77 sightlines),  implying that the gas is largely ionized toward every direction probed, $> 50$\% on average but as large as $>90$\% toward many sightlines, consistent with the high detection rate of \nii. The high occurrence of \ovi\ \citep[and substantial column densities, see][]{howk02,lehner07} implies that the HVC is also largely (possibly dominantly) highly ionized. 

4. Using the ratios of \siii/\sii\ and \siii/\feii, we conclude that there is some evidence of mild elemental depletion, which implies the presence of dust grains in these HVCs. 

5. From the sample where limits are excluded, we derive the mean  metallicity of the HVC  $[{\rm O/H}] = -0.51 \,^{+0.12}_{-0.16}$ (or ${\rm O/H}= 0.31 \pm 0.10$ solar, 44 sightlines). Many lower limits imply   $[{\rm O/H}] > -0.6$, and including those, the Kaplan-Meier estimator gives  $[{\rm O/H}] = -0.45$. There is no evidence of intrinsic variation in the abundance of oxygen, suggesting that these HVCs are part of a single complex that has the same origin. 

6. Based on stellar and nebular abundances, the present-day O abundances of the LMC and SMC are $ -0.4$ and $-0.7$ dex solar. Therefore, the HVC metallicity toward the LMC is more consistent with the LMC O abundance. Dust depletion may affect the O interstellar abundance by a small amount ($< 0.1$ dex). In that case, the (small) correction would increase the metallicity of the HVC, and hence $[{\rm O/H}]$ of the HVC complex  would converge to the present-day canonical LMC O abundance.

7. We do not find strong dependence in the relative or absolute abundances with any other quantities (e.g., galactic coordinates, \hi\ column density, or velocity). However, the average velocities of the HVCs in the LMCSR or LSR frames show a gradient with the right ascension (or galactic latitude). Interestingly, most of the supergiant shells believed to produce strong outflows in galaxies are situated at RA\,$>5.2^{\rm h}$ where the deviation velocities of the HVCs from the average velocities of the LMC disk  are the highest. 

8. We estimate that the mass of the neutral gas of the HVC complex is $\sim (0.5$--$1) \times 10^6 $ M$_\odot$ if the HVC complex is at 40--50 kpc. Including the photoionized and collisionally ionized gas, the total mass is likely $>0.5 \times 10^7$ M$_\odot$. This is comparable to or larger than the total mass of Complex C. 

9. All of our observational results are consistent with a generalized galactic outflow from the LMC as the origin of the HVC complex at LSR velocity 90--175 \km\ distributed across the LMC, including  the metallicity and kinematics of the HVCs as well as earlier findings from radio, optical and UV observations. This origin also naturally explains why the HVC is principally ionized, with both signatures of photoionization and collisional ionization. Dust grains are found in other known major galactic winds, and there is evidence of dust in this HVC complex as well.  However, we do not find any strong evidence of infalling material, therefore the HVC gas is likely to escape the LMC potential, suggesting that the LMC has a galactic wind. Ultimately  this HVC complex may be accreted by the MW and serve as additional fuel for future star formation within the MW. Based on these findings, it is possible that other large HVC complexes  found in the MW halo could be the result of outflows from nearby low mass galaxies (galaxies that could be now hidden by the MW disk).

\acknowledgments
We are grateful to Naomi McClure-Griffiths for making the GASS data available to us. Support for this research was made by NASA through grant NNX08AJ51G. This research has made use of the NASA Astrophysics Data System Abstract Service and the Centre de Donn\'ees de Strasbourg (CDS).

\clearpage
\begin{landscape}    
\LongTables
\begin{deluxetable}{lcccccccccccc}
\tablewidth{0pc}
\tablecaption{Summary of the Measurements for the HVCs toward the LMC \label{t-sum}}
\tabletypesize{\scriptsize}
\tablehead{
\colhead{Name} & \colhead{RA} & \colhead{DEC} & 
\colhead{$v$(\oi$)$} &\colhead{$\log N($\oi$)$} & 
\colhead{$v$(\feii$)$} &\colhead{$\log N($\feii$)$} & 
\colhead{$v$(\hi$)$} &\colhead{$\log N($\hi$)$} & 
\colhead{$[$\feii/\oi$]$} & \colhead{$[$\oi/\hi$]$} \\
\colhead{} & \colhead{($^{\rm h}$)} & \colhead{($\degr$)} & 
\colhead{(\km)} & \colhead{$[{\rm cm}^{-2}]$} &
\colhead{(\km)} & \colhead{$[{\rm cm}^{-2}]$} &
\colhead{(\km)} & \colhead{$[{\rm cm}^{-2}]$} &
\colhead{} &\colhead{}  }
\startdata
   Sk--67\D05     & $  4.8386 $  & $-67.6605 $ &	 \nodata	 &     \nodata  		     &      $  112.3 \pm   3.6 $ &  $	 13.79 \pm 0.06 	   $ &  	 \nodata	 &  $ <  18.63  	   $ &  	  \nodata		     &  	  \nodata		     \\ 
   Sk--68\D03     & $  4.8710 $  & $-68.4075 $ &	 \nodata	 &  $ <  14.37  		   $ &  	 \nodata	 &  $ <  13.19  		   $ &  	 \nodata	 &  $ <  18.47  	   $ &  	  \nodata		     &  	  \nodata		     \\
         BI12     & $  4.8836 $  & $-68.0257 $ &    $  140.2 \pm   1.9 $ &  $	 14.82 \pm 0.04 	   $ &      $  133.4 \pm   1.8 $ &  $	 13.62 \pm 0.05 	   $ &  	 \nodata	 &  $ <  18.50  	   $ &      $	  0.07 \pm\,^{0.06}_{0.07} $ &      $ >    -0.59		   $ \\
         BI13     & $  4.8851 $  & $-68.0564 $ &    $  138.9 \pm   0.8 $ &  $	 14.90 \pm 0.03 	   $ &      $  133.0 \pm   1.6 $ &  $	 13.65 \pm 0.05 	   $ &  	 \nodata	 &  $ <  18.49  	   $ &      $	  0.02 \pm 0.05 	   $ &      $ >    -0.48		   $ \\
   Sk--70\D13     & $  4.9003 $  & $-69.9965 $ &    $  142.9 \pm   1.9 $ &  $	 15.14 \pm 0.03 	   $ &      $  144.2 \pm   1.6 $ &  $	 13.89 \pm 0.06 	   $ &      $  145.8 \pm   1.1 $ &  $	 18.96 \pm 0.07 $ &	    $	  0.02 \pm\,^{0.07}_{0.08} $ &      $	 -0.55 \pm 0.08 	   $ \\
   Sk--65\D01     & $  4.9018 $  & $-65.5896 $ &    $  102.6 \pm   3.1 $ &  $	 15.05 \pm\,^{0.10}_{0.13} $ &      $  103.1 \pm   1.3 $ &  $	 13.93 \pm 0.06 	   $ &      $  100.4 \pm   2.8 $ &  $	 18.62 \pm 0.04 $ &	    $	  0.15 \pm\,^{0.11}_{0.16} $ &      $	 -0.30 \pm\,^{0.11}_{0.14} $ \\
   Sk--67\D18     & $  4.9208 $  & $-67.1901 $ &	 \nodata	 &  $ <  14.10  		   $ &  	 \nodata	 &     \nodata  		     &  	 \nodata	 &  $ <  18.47  	   $ &  	  \nodata		     &  	  \nodata		     \\
   Sk--67\D20     & $  4.9254 $  & $-67.5003 $ &	 \nodata	 &     \nodata  		     &  	 \nodata	 &  $ <  12.98  		   $ &  	 \nodata	 &  $ <  18.47  	   $ &  	  \nodata		     &  	  \nodata		     \\
   Sk--66\D18     & $  4.9333 $  & $-65.9750 $ &	 \nodata	 &     \nodata  		     &  	 \nodata	 &  $ <  13.29  		   $ &  	 \nodata	 &  $ <  18.47  	   $ &  	  \nodata		     &  	  \nodata		     \\
   Sk--66\D28     & $  4.9430 $  & $-66.4738 $ &    $  157.5 \pm   3.4 $ &  $	 14.50 \pm\,^{0.09}_{0.11} $ &      $  156.9 \pm   1.8 $ &  $	 13.64 \pm 0.05 	   $ &  	 \nodata	 &  $ <  18.48  	   $ &      $	  0.41 \pm\,^{0.10}_{0.13} $ &      $ >    -1.02		   $ \\
       BRRG39     & $  4.9447 $  & $-66.4112 $ &    $  167.6 \pm   1.8 $ &  $	 14.70 \pm\,^{0.06}_{0.07} $ &      $  158.3 \pm   3.0 $ &  $	 14.04 \pm 0.05 	   $ &  	 \nodata	 &  $ <  18.46  	   $ &      $	  0.61 \pm\,^{0.08}_{0.10} $ &      $ >    -0.73		   $ \\
       BRRG75     & $  4.9453 $  & $-66.4172 $ &	 \nodata	 &     \nodata  		     &      $  153.9 \pm   3.6 $ &  $	 13.61 \pm\,^{0.07}_{0.09} $ &      $  146.2 \pm   3.8 $ &  $	 18.62 \pm 0.15 $ &		  \nodata		     &  	  \nodata		     \\
     PGMW3073     & $  4.9454 $  & $-66.4150 $ &    $  155.9 \pm   2.3 $ &  $	 14.75 \pm\,^{0.08}_{0.09} $ &      $  159.9 \pm   6.4 $ &  $	 13.74 \pm\,^{0.12}_{0.16} $ &      $  146.2 \pm   3.8 $ &  $	 18.62 \pm 0.15 $ &	    $	  0.26 \pm\,^{0.14}_{0.20} $ &      $	 -0.60 \pm\,^{0.16}_{0.19} $ \\
     LH103102     & $  4.9459 $  & $-66.4128 $ &    $  147.3 \pm   3.5 $ &  $	 14.75 \pm\,^{0.08}_{0.09} $ &      $  153.5 \pm   2.5 $ &  $	 13.91 \pm 0.05 	   $ &      $  146.2 \pm   3.8 $ &  $	 18.62 \pm 0.15 $ &	    $	  0.43 \pm\,^{0.09}_{0.11} $ &      $	 -0.60 \pm\,^{0.16}_{0.19} $ \\
      BRRG140$^a$ & $  4.9463 $  & $-66.4129 $ &    $  146.4 \pm   2.0 $ &  $	 14.87 \pm 0.05 	   $ &      $  146.0 \pm   1.8 $ &  $	 13.85 \pm 0.08 	   $ &      $  146.2 \pm   3.8 $ &  $	 18.62 \pm 0.15 $ &	    $	  0.25 \pm\,^{0.09}_{0.11} $ &      $	 -0.48 \pm\,^{0.16}_{0.17} $ \\
    Sk--65\D2     & $  4.9476 $  & $-65.5190 $ &	 \nodata	 &  $ <  14.39  		   $ &  	 \nodata	 &     \nodata  		     &  	 \nodata	 &  $ <  18.47  	   $ &  	  \nodata		     &  	  \nodata		     \\
       N11032     & $  4.9484 $  & $-66.4044 $ &    $  149.7 \pm   2.1 $ &  $	 14.78 \pm 0.06 	   $ &      $  147.1 \pm   2.5 $ &  $	 13.78 \pm\,^{0.08}_{0.10} $ &      $  146.2 \pm   3.8 $ &  $	 18.62 \pm 0.15 $ &	    $	  0.27 \pm\,^{0.09}_{0.12} $ &      $	 -0.57 \pm\,^{0.16}_{0.17} $ \\
     LH103204     & $  4.9497 $  & $-66.4112 $ &    $  146.0 \pm   1.1 $ &  $	 14.80 \pm 0.04 	   $ &      $  146.9 \pm   1.9 $ &  $	 13.63 \pm\,^{0.07}_{0.08} $ &      $  146.2 \pm   3.8 $ &  $	 18.62 \pm 0.15 $ &	    $	  0.10 \pm\,^{0.08}_{0.09} $ &      $	 -0.55 \pm 0.15 	   $ \\
   SK--66\D33     & $  4.9497 $  & $-66.4106 $ &    $  148.3 \pm   2.2 $ &  $	 14.83 \pm 0.04 	   $ &      $  155.0 \pm   1.9 $ &  $	 13.84 \pm\,^{0.06}_{0.07} $ &      $  142.2 \pm   9.7 $ &  $	 18.62 \pm 0.14 $ &	    $	  0.28 \pm\,^{0.07}_{0.09} $ &      $	 -0.52 \pm 0.14 	   $ \\
         BI42     & $  4.9502 $  & $-66.4070 $ &    $  141.4 \pm   1.4 $ &  $	 14.87 \pm 0.06 	   $ &      $  147.5 \pm   2.3 $ &  $	 13.84 \pm\,^{0.07}_{0.09} $ &      $  142.2 \pm   9.7 $ &  $	 18.62 \pm 0.14 $ &	    $	  0.24 \pm\,^{0.09}_{0.12} $ &      $	 -0.48 \pm\,^{0.15}_{0.16} $ \\
   Sk--66\D35     & $  4.9513 $  & $-66.5773 $ &	 \nodata	 &     \nodata  		     &  	 \nodata	 &  $ <  13.44  		   $ &  	 \nodata	 &  $ <  18.47  	   $ &  	  \nodata		     &  	  \nodata		     \\
   Sk--69\D50     & $  4.9542 $  & $-69.3390 $ &    $  147.9 \pm   0.9 $ &  $	 15.42 \pm 0.02 	   $ &      $  145.3 \pm   1.2 $ &  $	 14.39 \pm 0.03 	   $ &      $  146.9 \pm   3.8 $ &  $	 19.23 \pm 0.10 $ &	    $	  0.24 \pm 0.04 	   $ &      $	 -0.54 \pm 0.10 	   $ \\
       HV2241     & $  4.9544 $  & $-66.5651 $ &    $  148.6 \pm   2.5 $ &  $	 14.49 \pm 0.05 	   $ &      $  148.5 \pm   2.4 $ &  $	 13.37 \pm\,^{0.07}_{0.08} $ &  	 \nodata	 &  $ <  18.53  	   $ &      $	  0.15 \pm\,^{0.08}_{0.10} $ &      $ >    -0.97		   $ \\
       IC2116     & $  4.9545 $  & $-66.3892 $ &    $  152.0 \pm   6.0 $ &  $	 14.79 \pm\,^{0.15}_{0.24} $ &      $  146.2 \pm   5.2 $ &  $	 13.89 \pm\,^{0.16}_{0.27} $ &      $  148.8 \pm   3.5 $ &  $	 18.69 \pm 0.09 $ &	    $	  0.36 \pm\,^{0.21}_{0.42} $ &      $	 -0.63 \pm\,^{0.17}_{0.27} $ \\
   Sk--68\D15     & $  4.9567 $  & $-68.3993 $ &    $  144.0 \pm   1.6 $ &  $	 14.60 \pm\,^{0.12}_{0.16} $ &      $  140.6 \pm   4.4 $ &  $	 13.76 \pm\,^{0.12}_{0.17} $ &      $  144.5 \pm   9.9 $ &  $	 18.22 \pm 0.15 $ &	    $	  0.43 \pm\,^{0.16}_{0.26} $ &      $	 -0.35 \pm\,^{0.18}_{0.24} $ \\
   Sk--67\D22     & $  4.9577 $  & $-67.6508 $ &	 \nodata	 &     \nodata  		     &  	 \nodata	 &  $ <  13.17  		   $ &  	 \nodata	 &  $ <  18.47  	   $ &  	  \nodata		     &  	  \nodata		     \\
   Sk--68\D16     & $  4.9605 $  & $-68.4100 $ &    $  135.3 \pm   2.9 $ &  $	 14.75 \pm\,^{0.07}_{0.09} $ &      $  138.0 \pm   1.5 $ &  $	 13.75 \pm\,^{0.06}_{0.08} $ &      $  139.8 \pm   8.0 $ &  $	 18.45 \pm 0.15 $ &	    $	  0.27 \pm\,^{0.10}_{0.12} $ &      $	 -0.43 \pm\,^{0.16}_{0.18} $ \\
   Sk--67\D28     & $  4.9776 $  & $-67.1886 $ &	 \nodata	 &  $ <  14.28  		   $ &  	 \nodata	 &  $ <  13.19  		   $ &  	 \nodata	 &  $ <  18.47  	   $ &  	  \nodata		     &  	  \nodata		     \\
   Sk--67\D32     & $  4.9977 $  & $-67.9489 $ &	 \nodata	 &  $ <  14.03  		   $ &  	 \nodata	 &  $ <  13.23  		   $ &  	 \nodata	 &  $ <  18.47  	   $ &  	  \nodata		     &  	  \nodata		     \\
   Sk--70\D32     & $  5.0029 $  & $-70.1860 $ &    $  137.6 \pm   2.0 $ &  $	 15.35 \pm 0.04 	   $ &      $  143.3 \pm   1.1 $ &  $	 14.24 \pm 0.03 	   $ &      $  142.8 \pm   3.7 $ &  $	 19.00 \pm 0.15 $ &	    $	  0.16 \pm 0.05 	   $ &      $	 -0.38 \pm 0.15 	   $ \\
   Sk--65\D21     & $  5.0229 $  & $-65.6967 $ &	 \nodata	 &  $ <  14.28  		   $ &      $  124.8 \pm   3.6 $ &  $	 13.73 \pm 0.05 	   $ &  	 \nodata	 &  $ <  18.59  	   $ &      $ >    0.66 		   $ &  	  \nodata		     \\
   Sk--65\D22$^a$ & $  5.0231 $  & $-65.8760 $ &    $  137.3 \pm   4.6 $ &  $	 14.34 \pm 0.05 	   $ &      $  136.7 \pm   1.8 $ &  $	 13.57 \pm 0.02 	   $ &      $  137.5 \pm   4.8 $ &  $	 18.45 \pm 0.17 $ &	    $	  0.50 \pm\,^{0.11}_{0.16} $ &      $	 -0.84 \pm\,^{0.19}_{0.25} $ \\
   Sk--68\D26     & $  5.0256 $  & $-68.1786 $ &	 \nodata	 &     \nodata  		     &  	 \nodata	 &  $ <  13.62  		   $ &  	 \nodata	 &  $ <  18.47  	   $ &  	  \nodata		     &  	  \nodata		     \\
       HV2274     & $  5.0447 $  & $-68.4059 $ &	 \nodata	 &  $ <  14.44  		   $ &  	 \nodata	 &  $ <  13.16  		   $ &  	 \nodata	 &  $ <  18.47  	   $ &  	  \nodata		     &  	  \nodata		     \\
   Sk--66\D51     & $  5.0528 $  & $-66.6817 $ &	 \nodata	 &     \nodata  		     &  	 \nodata	 &  $ <  13.16  		   $ &  	 \nodata	 &  $ <  18.47  	   $ &  	  \nodata		     &  	  \nodata		     \\
   Sk--69\D59     & $  5.0536 $  & $-69.0270 $ &    $  175.9 \pm   0.6 $ &  $	 15.06 \pm 0.02 	   $ &      $  173.4 \pm   1.3 $ &  $	 13.79 \pm 0.03 	   $ &      $  172.4 \pm   2.5 $ &  $	 18.89 \pm 0.11 $ &	    $	  0.00 \pm 0.04 	   $ &      $	 -0.56 \pm 0.11 	   $ \\
   Sk--67\D38     & $  5.0583 $  & $-67.8737 $ &    $  125.2 \pm   2.9 $ &  $	 14.23 \pm\,^{0.10}_{0.13} $ &      $  128.1 \pm   2.0 $ &  $	 13.47 \pm\,^{0.08}_{0.09} $ &  	 \nodata	 &  $ <  18.44  	   $ &      $	  0.51 \pm\,^{0.12}_{0.16} $ &      $ >    -1.29		   $ \\
   NGC1818-D1     & $  5.0757 $  & $-66.4142 $ &	 \nodata	 &  $ <  14.60  		   $ &      $  108.3 \pm   3.8 $ &  $	 13.83 \pm\,^{0.09}_{0.12} $ &  	 \nodata	 &  $ <  18.55  	   $ &      $ >    0.39 		   $ &  	  \nodata		     \\
   Sk--70\D60     & $  5.0780 $  & $-70.2596 $ &    $  123.3 \pm   1.1 $ &  $	 14.64 \pm\,^{0.07}_{0.08} $ &      $  128.7 \pm   1.7 $ &  $	 13.79 \pm\,^{0.06}_{0.08} $ &  	 \nodata	 &  $ <  18.45  	   $ &      $	  0.41 \pm\,^{0.09}_{0.12} $ &      $ >    -0.80		   $ \\
   Sk--70\D69     & $  5.0885 $  & $-70.4305 $ &    $  120.5 \pm   2.9 $ &  $	 14.71 \pm\,^{0.07}_{0.08} $ &      $  108.7 \pm   1.7 $ &  $	 13.63 \pm\,^{0.07}_{0.08} $ &  	 \nodata	 &  $ <  18.48  	   $ &      $	  0.20 \pm\,^{0.10}_{0.12} $ &      $ >    -0.77		   $ \\
   Sk--68\D41     & $  5.0909 $  & $-68.1674 $ &	 \nodata	 &  $ <  13.92  		   $ &  	 \nodata	 &     \nodata  		     &  	 \nodata	 &  $ <  18.47  	   $ &  	  \nodata		     &  	  \nodata		     \\
   Sk--70\D78     & $  5.1045 $  & $-70.4932 $ &    $	97.2 \pm   4.4 $ &  $	 14.27 \pm\,^{0.15}_{0.23} $ &  	 \nodata	 &     \nodata  		     &  	 \nodata	 &  $ <  18.44  	   $ &  	  \nodata		     &      $ >    -1.45		   $ \\
   Sk--70\D79     & $  5.1104 $  & $-70.4902 $ &	 \nodata	 &  $ <  14.55  		   $ &  	 \nodata	 &     \nodata  		     &  	 \nodata	 &  $ <  18.47  	   $ &  	  \nodata		     &  	  \nodata		     \\
   Sk--67\D46     & $  5.1171 $  & $-67.6249 $ &	 \nodata	 &     \nodata  		     &  	 \nodata	 &  $ <  13.69  		   $ &  	 \nodata	 &  $ <  18.47  	   $ &  	  \nodata		     &  	  \nodata		     \\
   Sk--68\D52     & $  5.1224 $  & $-68.5360 $ &	 \nodata	 &  $ <  14.26  		   $ &  	 \nodata	 &  $ <  13.12  		   $ &  	 \nodata	 &  $ <  18.47  	   $ &  	  \nodata		     &  	  \nodata		     \\
   Sk--71\D08     & $  5.1232 $  & $-71.1985 $ &	 \nodata	 &     \nodata  		     &      $  140.2 \pm   2.8 $ &  $	 13.55 \pm\,^{0.10}_{0.13} $ &  	 \nodata	 &  $ <  18.41  	   $ &  	  \nodata		     &  	  \nodata		     \\
 MACHO79-4779     & $  5.1581 $  & $-68.9175 $ &	 \nodata	 &  $ <  14.58  		   $ &  	 \nodata	 &  $ <  13.32  		   $ &  	 \nodata	 &  $ <  18.47  	   $ &  	  \nodata		     &  	  \nodata		     \\
   Sk--68\D57     & $  5.1613 $  & $-68.8903 $ &	 \nodata	 &  $ <  14.37  		   $ &  	 \nodata	 &  $ <  13.36  		   $ &  	 \nodata	 &  $ <  18.47  	   $ &  	  \nodata		     &  	  \nodata		     \\
   Sk--69\D79     & $  5.2317 $  & $-69.5300 $ &	 \nodata	 &  $ <  14.29  		   $ &  	 \nodata	 &  $ <  13.16  		   $ &  	 \nodata	 &  $ <  18.47  	   $ &  	  \nodata		     &  	  \nodata		     \\
   Sk--67\D69     & $  5.2389 $  & $-67.1343 $ &	 \nodata	 &     \nodata  		     &  	 \nodata	 &  $ <  13.12  		   $ &  	 \nodata	 &  $ <  18.47  	   $ &  	  \nodata		     &  	  \nodata		     \\
   Sk--70\D85     & $  5.2849 $  & $-70.3231 $ &	 \nodata	 &     \nodata  		     &      $	93.3 \pm   5.2 $ &  $	 13.62 \pm\,^{0.10}_{0.12} $ &  	 \nodata	 &  $ <  18.57  	   $ &  	  \nodata		     &  	  \nodata		     \\
   Sk--69\D95     & $  5.3053 $  & $-69.1946 $ &	 \nodata	 &  $ <  14.36  		   $ &  	 \nodata	 &  $ <  13.36  		   $ &  	 \nodata	 &  $ <  18.47  	   $ &  	  \nodata		     &  	  \nodata		     \\
        BI128     & $  5.3055 $  & $-65.8206 $ &	 \nodata	 &  $ <  14.49  		   $ &      $  130.8 \pm   2.5 $ &  $	 14.04 \pm 0.05 	   $ &  	 \nodata	 &  $ <  18.61  	   $ &      $ >    0.77 		   $ &  	  \nodata		     \\
  Sk--69\D104     & $  5.3165 $  & $-69.2152 $ &	 \nodata	 &     \nodata  		     &      $  118.3 \pm   2.7 $ &  $	 13.58 \pm\,^{0.07}_{0.09} $ &  	 \nodata	 &  $ <  18.57  	   $ &  	  \nodata		     &  	  \nodata		     \\
       BREY22     & $  5.3212 $  & $-69.6554 $ &    $  146.2 \pm   1.1 $ &  $	 15.32 \pm 0.10 	   $ &      $  143.9 \pm   0.4 $ &  $	 14.39 \pm 0.01 	   $ &  	 \nodata	 &  $ <  18.57  	   $ &      $	  0.34 \pm 0.10 	   $ &      $ >    -0.27		   $ \\
   Sk--67\D76     & $  5.3349 $  & $-67.3525 $ &    $  110.8 \pm   2.9 $ &  $	 14.93 \pm 0.06 	   $ &      $  117.3 \pm   2.3 $ &  $	 13.93 \pm 0.05 	   $ &  	 \nodata	 &  $ <  18.59  	   $ &      $	  0.26 \pm\,^{0.07}_{0.09} $ &      $ >    -0.61		   $ \\
   Sk--65\D44     & $  5.3383 $  & $-65.4036 $ &	 \nodata	 &  $ <  14.18  		   $ &  	 \nodata	 &  $ <  13.13  		   $ &  	 \nodata	 &  $ <  18.47  	   $ &  	  \nodata		     &  	  \nodata		     \\
   Sk--65\D47     & $  5.3485 $  & $-65.4550 $ &    $  144.1 \pm   4.9 $ &  $	 14.04 \pm\,^{0.10}_{0.12} $ &  	 \nodata	 &  $ <  12.80  		   $ &  	 \nodata	 &  $ <  18.47  	   $ &      $ <    0.13 		   $ &      $ >    -1.51		   $ \\
       Brey24     & $  5.3660 $  & $-65.8164 $ &	 \nodata	 &  $ <  13.88  		   $ &      $  113.9 \pm   3.6 $ &  $	 13.17 \pm\,^{0.09}_{0.11} $ &  	 \nodata	 &  $ <  18.42  	   $ &      $ >    0.45 		   $ &  	  \nodata		     \\
     LH47-338     & $  5.3706 $  & $-67.9768 $ &	 \nodata	 &     \nodata  		     &  	 \nodata	 &  $ <  13.15  		   $ &  	 \nodata	 &  $ <  18.47  	   $ &  	  \nodata		     &  	  \nodata		     \\
   Sk--71\D19     & $  5.3711 $  & $-71.3611 $ &    $  146.2 \pm   2.7 $ &  $	 15.11 \pm 0.06 	   $ &      $  142.6 \pm   1.2 $ &  $	 14.19 \pm 0.03 	   $ &  	 \nodata	 &  $ <  18.57  	   $ &      $	  0.34 \pm\,^{0.06}_{0.07} $ &      $ >    -0.41		   $ \\
   Sk--71\D21     & $  5.3729 $  & $-71.5994 $ &    $  136.6 \pm   3.5 $ &  $	 14.99 \pm\,^{0.07}_{0.09} $ &      $  143.4 \pm   1.7 $ &  $	 14.01 \pm 0.05 	   $ &  	 \nodata	 &  $ <  18.55  	   $ &      $	  0.28 \pm\,^{0.08}_{0.11} $ &      $ >    -0.56		   $ \\
   Sk--68\D73     & $  5.3833 $  & $-68.0296 $ &	 \nodata	 &     \nodata  		     &      $  126.3 \pm   2.9 $ &  $	 13.85 \pm 0.06 	   $ &  	 \nodata	 &  $ <  18.54  	   $ &  	  \nodata		     &  	  \nodata		     \\
   Sk--71\D26     & $  5.3861 $  & $-71.3474 $ &    $  135.6 \pm   1.0 $ &  $	 14.69 \pm 0.04 	   $ &      $  134.9 \pm   1.5 $ &  $	 14.07 \pm 0.04 	   $ &  	 \nodata	 &  $ <  18.56  	   $ &      $	  0.65 \pm 0.05 	   $ &      $ >    -0.79		   $ \\
   Sk--68\D75     & $  5.3913 $  & $-68.2063 $ &    $  114.7 \pm   2.1 $ &  $	 14.96 \pm 0.05 	   $ &      $  119.5 \pm   2.2 $ &  $	 13.96 \pm\,^{0.07}_{0.08} $ &      $  120.5 \pm   5.6 $ &  $	 18.86 \pm 0.09 $ &	    $	  0.27 \pm\,^{0.08}_{0.10} $ &      $	 -0.63 \pm 0.10 	   $ \\
   Sk--66\D78     & $  5.3918 $  & $-66.7032 $ &	 \nodata	 &  $ <  14.47  		   $ &  	 \nodata	 &  $ <  13.56  		   $ &  	 \nodata	 &  $ <  18.47  	   $ &  	  \nodata		     &  	  \nodata		     \\
  Sk--69\D124     & $  5.4218 $  & $-69.0531 $ &    $  131.7 \pm   2.0 $ &  $	 14.90 \pm 0.03 	   $ &      $  128.1 \pm   1.0 $ &  $	 13.89 \pm 0.03 	   $ &      $  126.8 \pm   1.9 $ &  $	 18.83 \pm 0.10 $ &	    $	  0.26 \pm 0.04 	   $ &      $	 -0.66 \pm 0.10 	   $ \\
  Sk--67\D101$^a$ & $  5.4323 $  & $-67.5080 $ &    $  110.6 \pm   1.2 $ &  $	 15.13 \pm 0.02 	   $ &      $  110.6 \pm   0.9 $ &  $	 14.15 \pm 0.02 	   $ &      $  114.6 \pm   2.8 $ &  $	 18.79 \pm 0.12 $ &	    $	  0.29 \pm 0.03 	   $ &      $	 -0.39 \pm 0.12 	   $ \\
  Sk--67\D104$^a$ & $  5.4345 $  & $-67.4990 $ &    $  113.2 \pm   0.8 $ &  $	 15.21 \pm 0.02 	   $ &      $  114.1 \pm   0.6 $ &  $	 14.24 \pm 0.01 	   $ &      $  114.6 \pm   2.8 $ &  $	 18.79 \pm 0.12 $ &	    $	  0.30 \pm 0.03 	   $ &      $	 -0.31 \pm 0.12 	   $ \\
  Sk--67\D105     & $  5.4351 $  & $-67.1827 $ &    $  118.9 \pm   2.3 $ &  $	 15.00 \pm 0.03 	   $ &      $  116.0 \pm   1.7 $ &  $	 13.89 \pm 0.03 	   $ &      $  116.2 \pm   5.6 $ &  $	 18.80 \pm 0.12 $ &	    $	  0.16 \pm 0.04 	   $ &      $	 -0.53 \pm 0.12 	   $ \\
  Sk--67\D106$^a$ & $  5.4376 $  & $-67.4995 $ &    $  110.7 \pm   0.6 $ &  $	 15.20 \pm 0.02 	   $ &      $  112.2 \pm   1.1 $ &  $	 14.22 \pm 0.02 	   $ &      $  114.2 \pm   2.8 $ &  $	 18.81 \pm 0.12 $ &	    $	  0.29 \pm 0.02 	   $ &      $	 -0.34 \pm 0.12 	   $ \\
  Sk--67\D107$^a$ & $  5.4391 $  & $-67.4987 $ &    $  111.7 \pm   0.8 $ &  $	 15.18 \pm 0.02 	   $ &      $  115.4 \pm   1.2 $ &  $	 14.25 \pm 0.02 	   $ &      $  114.2 \pm   2.8 $ &  $	 18.81 \pm 0.12 $ &	    $	  0.34 \pm 0.03 	   $ &      $	 -0.36 \pm 0.12 	   $ \\
     LH54-425     & $  5.4401 $  & $-67.5048 $ &    $  113.5 \pm   0.6 $ &  $	 15.16 \pm 0.01 	   $ &      $  105.5 \pm   0.6 $ &  $	 14.03 \pm 0.02 	   $ &      $  114.2 \pm   2.8 $ &  $	 18.81 \pm 0.12 $ &	    $	  0.14 \pm 0.02 	   $ &      $	 -0.38 \pm 0.12 	   $ \\
  Sk--67\D108     & $  5.4407 $  & $-67.6223 $ &    $  107.5 \pm   0.9 $ &  $	 15.08 \pm 0.03 	   $ &      $	98.6 \pm   1.0 $ &  $	 14.14 \pm 0.03 	   $ &      $  116.3 \pm   1.2 $ &  $	 19.00 \pm 0.12 $ &	    $	  0.33 \pm 0.04 	   $ &      $	 -0.65 \pm 0.12 	   $ \\
   Sk--68\D80     & $  5.4418 $  & $-68.8404 $ &    $  124.7 \pm   1.2 $ &  $	 15.15 \pm 0.03 	   $ &      $  115.7 \pm   0.4 $ &  $	 14.15 \pm 0.01 	   $ &      $  126.2 \pm   1.6 $ &  $	 19.07 \pm 0.07 $ &	    $	  0.27 \pm 0.03 	   $ &      $	 -0.65 \pm 0.08 	   $ \\
   Sk--68\D82     & $  5.4459 $  & $-68.8313 $ &    $  121.0 \pm   0.4 $ &  $	 15.18 \pm 0.01 	   $ &      $  108.9 \pm   2.2 $ &  $	 14.15 \pm 0.05 	   $ &      $  126.2 \pm   1.6 $ &  $	 19.07 \pm 0.07 $ &	    $	  0.24 \pm 0.05 	   $ &      $	 -0.62 \pm 0.07 	   $ \\
        BI170     & $  5.4466 $  & $-69.1032 $ &	 \nodata	 &     \nodata  		     &      $  104.5 \pm   2.5 $ &  $	 13.97 \pm 0.06 	   $ &      $  104.0 \pm   3.5 $ &  $	 18.99 \pm 0.12 $ &		  \nodata		     &  	  \nodata		     \\
  Sk--67\D111     & $  5.4467 $  & $-67.4916 $ &    $  108.1 \pm   0.4 $ &  $	 15.08 \pm 0.01 	   $ &      $  108.2 \pm   0.4 $ &  $	 14.09 \pm 0.01 	   $ &      $  112.9 \pm   2.1 $ &  $	 18.84 \pm 0.08 $ &	    $	  0.28 \pm 0.01 	   $ &      $	 -0.49 \pm 0.08 	   $ \\
        BI173     & $  5.4528 $  & $-69.1323 $ &    $  116.6 \pm   0.7 $ &  $	 15.05 \pm 0.02 	   $ &      $  109.2 \pm   0.9 $ &  $	 14.13 \pm 0.02 	   $ &      $  117.9 \pm   2.4 $ &  $	 18.99 \pm 0.06 $ &	    $	  0.35 \pm 0.03 	   $ &      $	 -0.67 \pm 0.06 	   $ \\
   Sk--66\D97     & $  5.4551 $  & $-66.3687 $ &	 \nodata	 &  $ <  14.53  		   $ &  	 \nodata	 &  $ <  13.50  		   $ &  	 \nodata	 &  $ <  18.47  	   $ &  	  \nodata		     &  	  \nodata		     \\
       HV2243     & $  5.4576 $  & $-67.1987 $ &    $  105.1 \pm   1.9 $ &  $	 14.85 \pm\,^{0.07}_{0.08} $ &      $  106.4 \pm   1.9 $ &  $	 13.98 \pm 0.06 	   $ &  	 \nodata	 &  $ <  18.45  	   $ &      $	  0.40 \pm\,^{0.09}_{0.11} $ &      $ >    -0.58		   $ \\
  Sk--67\D118     & $  5.4593 $  & $-67.2917 $ &    $  114.5 \pm   2.6 $ &  $	 14.82 \pm 0.06 	   $ &      $  112.3 \pm   3.0 $ &  $	 14.03 \pm\,^{0.07}_{0.08} $ &      $  119.4 \pm  10.1 $ &  $	 18.55 \pm 0.15 $ &	    $	  0.48 \pm\,^{0.09}_{0.11} $ &      $	 -0.46 \pm\,^{0.16}_{0.17} $ \\
   Sk--70\D91     & $  5.4594 $  & $-70.6134 $ &    $  116.0 \pm   3.2 $ &  $	 14.72 \pm\,^{0.11}_{0.16} $ &      $  112.5 \pm   0.8 $ &  $	 13.86 \pm 0.03 	   $ &  	 \nodata	 &  $ <  18.60  	   $ &      $	  0.41 \pm\,^{0.12}_{0.16} $ &      $ >    -1.02		   $ \\
  Sk--67\D119     & $  5.4613 $  & $-67.3029 $ &    $  112.9 \pm   2.1 $ &  $	 14.98 \pm 0.06 	   $ &      $  109.8 \pm   2.2 $ &  $	 13.96 \pm\,^{0.07}_{0.08} $ &      $  114.0 \pm   7.6 $ &  $	 18.55 \pm 0.12 $ &	    $	  0.25 \pm\,^{0.09}_{0.11} $ &      $	 -0.30 \pm\,^{0.13}_{0.14} $ \\
  Sk--66\D100     & $  5.4627 $  & $-66.9208 $ &    $  107.3 \pm   3.1 $ &  $	 14.59 \pm 0.06 	   $ &      $  109.6 \pm   1.4 $ &  $	 13.74 \pm 0.04 	   $ &  	 \nodata	 &  $ <  18.48  	   $ &      $	  0.42 \pm\,^{0.07}_{0.09} $ &      $ >    -0.85		   $ \\
  Sk--69\D142     & $  5.4647 $  & $-68.9857 $ &    $  121.5 \pm   2.2 $ &  $	 15.26 \pm 0.05 	   $ &      $  120.9 \pm   0.9 $ &  $	 14.22 \pm 0.03 	   $ &      $  118.7 \pm   2.5 $ &  $	 19.11 \pm 0.06 $ &	    $	  0.23 \pm 0.06 	   $ &      $	 -0.58 \pm 0.08 	   $ \\
  Sk--69\D147     & $  5.4730 $  & $-69.1423 $ &    $  119.7 \pm   3.4 $ &  $	 15.20 \pm\,^{0.08}_{0.10} $ &      $  110.6 \pm   1.6 $ &  $	 13.99 \pm 0.05 	   $ &      $  119.3 \pm   5.2 $ &  $	 19.07 \pm 0.09 $ &	    $	  0.05 \pm\,^{0.09}_{0.11} $ &      $	 -0.60 \pm\,^{0.12}_{0.14} $ \\
   Sk--65\D63     & $  5.4777 $  & $-65.6501 $ &    $  121.8 \pm   0.9 $ &  $	 14.63 \pm 0.04 	   $ &      $  123.6 \pm   1.3 $ &  $	 13.88 \pm 0.06 	   $ &  	 \nodata	 &  $ <  18.41  	   $ &      $	  0.52 \pm\,^{0.07}_{0.09} $ &      $ >    -0.70		   $ \\
  Sk--66\D106     & $  5.4836 $  & $-66.6411 $ &	 \nodata	 &  $ <  14.63  		   $ &  	 \nodata	 &     \nodata  		     &  	 \nodata	 &  $ <  18.47  	   $ &  	  \nodata		     &  	  \nodata		     \\
        HV982     & $  5.4979 $  & $-69.1561 $ &    $  112.1 \pm   1.9 $ &  $	 15.19 \pm\,^{0.09}_{0.11} $ &      $  112.9 \pm   2.0 $ &  $	 14.17 \pm\,^{0.11}_{0.14} $ &      $  117.1 \pm   2.5 $ &  $	 19.11 \pm 0.05 $ &	    $	  0.25 \pm\,^{0.13}_{0.19} $ &      $	 -0.65 \pm\,^{0.10}_{0.12} $ \\
   Sk--70\D97     & $  5.5032 $  & $-70.8617 $ &    $  111.3 \pm   2.6 $ &  $	 14.82 \pm 0.04 	   $ &      $  110.0 \pm   1.5 $ &  $	 13.79 \pm 0.06 	   $ &      $  112.3 \pm   1.4 $ &  $	 18.78 \pm 0.10 $ &	    $	  0.24 \pm\,^{0.07}_{0.09} $ &      $	 -0.69 \pm 0.11 	   $ \\
  Sk--67\D144     & $  5.5034 $  & $-67.4357 $ &    $  108.9 \pm   1.1 $ &  $	 15.00 \pm 0.03 	   $ &      $  103.7 \pm   1.6 $ &  $	 14.22 \pm 0.04 	   $ &  	 \nodata	 &  $ <  18.58  	   $ &      $	  0.49 \pm 0.04 	   $ &      $ >    -0.46		   $ \\
        BI184     & $  5.5085 $  & $-71.0420 $ &    $  106.0 \pm   1.9 $ &  $	 15.06 \pm 0.06 	   $ &      $  109.7 \pm   1.2 $ &  $	 14.11 \pm 0.05 	   $ &      $  105.2 \pm   6.0 $ &  $	 18.93 \pm 0.05 $ &	    $	  0.32 \pm\,^{0.08}_{0.09} $ &      $	 -0.60 \pm\,^{0.07}_{0.09} $ \\
  Sk--67\D150     & $  5.5088 $  & $-67.0148 $ &	 \nodata	 &  $ <  14.39  		   $ &  	 \nodata	 &     \nodata  		     &  	 \nodata	 &  $ <  18.47  	   $ &  	  \nodata		     &  	  \nodata		     \\
  NGC2004-B15     & $  5.5102 $  & $-67.2951 $ &    $  106.8 \pm   8.1 $ &  $	 14.87 \pm\,^{0.11}_{0.15} $ &      $  110.5 \pm   2.7 $ &  $	 14.28 \pm\,^{0.08}_{0.09} $ &  	 \nodata	 &  $ <  18.39  	   $ &      $	  0.68 \pm\,^{0.13}_{0.18} $ &      $ >    -0.64		   $ \\
   Sk--71\D38     & $  5.5108 $  & $-71.0300 $ &    $  117.6 \pm   1.1 $ &  $	 14.90 \pm 0.06 	   $ &      $  115.6 \pm   1.1 $ &  $	 13.69 \pm 0.05 	   $ &      $  119.2 \pm   4.7 $ &  $	 18.88 \pm 0.07 $ &	    $	  0.06 \pm\,^{0.08}_{0.09} $ &      $	 -0.71 \pm\,^{0.09}_{0.10} $ \\
   Sk--71\D41     & $  5.5112 $  & $-71.0937 $ &    $  114.2 \pm   0.9 $ &  $	 15.09 \pm 0.05 	   $ &      $  122.8 \pm   1.3 $ &  $	 14.11 \pm\,^{0.07}_{0.09} $ &      $  117.8 \pm   1.5 $ &  $	 18.96 \pm 0.06 $ &	    $	  0.29 \pm\,^{0.09}_{0.11} $ &      $	 -0.60 \pm\,^{0.08}_{0.09} $ \\
  NGC2004-B30     & $  5.5132 $  & $-67.2897 $ &    $	95.8 \pm   2.2 $ &  $	 14.92 \pm\,^{0.08}_{0.10} $ &  	 \nodata	 &     \nodata  		     &  	 \nodata	 &  $ <  18.45  	   $ &  	  \nodata		     &      $ >    -0.56		   $ \\
  Sk--68\D111     & $  5.5169 $  & $-68.8991 $ &    $  109.6 \pm   3.8 $ &  $	 15.15 \pm\,^{0.11}_{0.15} $ &      $  102.4 \pm   1.4 $ &  $	 14.35 \pm 0.05 	   $ &      $  114.5 \pm   3.8 $ &  $	 19.01 \pm 0.06 $ &	    $	  0.47 \pm\,^{0.12}_{0.17} $ &      $	 -0.59 \pm\,^{0.12}_{0.17} $ \\
   Sk--71\D45$^a$ & $  5.5210 $  & $-71.0691 $ &    $  125.0 \pm   2.9 $ &  $	 14.99 \pm 0.02 	   $ &      $  120.5 \pm   1.5 $ &  $	 13.83 \pm 0.02 	   $ &      $  117.7 \pm   3.1 $ &  $	 18.93 \pm 0.08 $ &	    $	  0.11 \pm 0.03 	   $ &      $	 -0.67 \pm 0.08 	   $ \\
  Sk--69\D175     & $  5.5238 $  & $-69.0940 $ &    $  114.9 \pm   1.2 $ &  $	 15.25 \pm 0.04 	   $ &      $  117.2 \pm   1.1 $ &  $	 14.26 \pm 0.03 	   $ &      $  116.4 \pm   4.5 $ &  $	 19.18 \pm 0.10 $ &	    $	  0.28 \pm 0.05 	   $ &      $	 -0.66 \pm 0.11 	   $ \\
  Sk--67\D161     & $  5.5258 $  & $-67.6796 $ &	 \nodata	 &  $ <  14.19  		   $ &  	 \nodata	 &  $ <  13.19  		   $ &  	 \nodata	 &  $ <  18.47  	   $ &  	  \nodata		     &  	  \nodata		     \\
  Sk--67\D166     & $  5.5290 $  & $-67.6335 $ &    $  121.4 \pm   0.7 $ &  $	 15.06 \pm 0.01 	   $ &      $  119.1 \pm   0.3 $ &  $	 14.14 \pm 0.01 	   $ &  	 \nodata	 &  $ <  18.66  	   $ &      $	  0.35 \pm 0.01 	   $ &      $ >    -0.45		   $ \\
  Sk--67\D169     & $  5.5310 $  & $-67.0395 $ &	 \nodata	 &  $ <  14.11  		   $ &  	 \nodata	 &     \nodata  		     &  	 \nodata	 &  $ <  18.47  	   $ &  	  \nodata		     &  	  \nodata		     \\
  Sk--67\D167     & $  5.5311 $  & $-67.6614 $ &    $  128.9 \pm   1.7 $ &  $	 14.90 \pm 0.03 	   $ &      $  118.7 \pm   1.3 $ &  $	 14.03 \pm 0.02 	   $ &  	 \nodata	 &  $ <  18.65  	   $ &      $	  0.40 \pm 0.04 	   $ &      $ >    -0.64		   $ \\
  Sk--67\D168     & $  5.5311 $  & $-67.5724 $ &    $  110.5 \pm   2.6 $ &  $	 15.00 \pm\,^{0.06}_{0.07} $ &      $  114.5 \pm   1.3 $ &  $	 14.13 \pm 0.03 	   $ &  	 \nodata	 &  $ <  18.55  	   $ &      $	  0.40 \pm\,^{0.07}_{0.08} $ &      $ >    -0.53		   $ \\
  Sk--67\D174     & $  5.5365 $  & $-67.6878 $ &    $  102.9 \pm   1.8 $ &  $	 14.90 \pm\,^{0.06}_{0.07} $ &      $  108.6 \pm   1.5 $ &  $	 14.13 \pm 0.04 	   $ &  	 \nodata	 &  $ <  18.50  	   $ &      $	  0.50 \pm\,^{0.08}_{0.09} $ &      $ >    -0.58		   $ \\
        BI196     & $  5.5387 $  & $-67.8306 $ &    $  107.7 \pm   2.4 $ &  $	 14.85 \pm 0.05 	   $ &      $  114.6 \pm   1.8 $ &  $	 14.07 \pm 0.05 	   $ &  	 \nodata	 &  $ <  18.56  	   $ &      $	  0.49 \pm\,^{0.07}_{0.08} $ &      $ >    -0.66		   $ \\
   4U0532-664$^a$ & $  5.5471 $  & $-66.3704 $ &    $  100.7 \pm   2.7 $ &  $	 14.63 \pm 0.03 	   $ &      $	98.0 \pm   1.3 $ &  $	 13.98 \pm 0.02 	   $ &  	 \nodata	 &  $ <  18.52  	   $ &      $	  0.62 \pm 0.05 	   $ &      $ >    -0.78		   $ \\
  Sk--67\D184     & $  5.5531 $  & $-67.7125 $ &    $  133.3 \pm   4.9 $ &  $	 14.96 \pm\,^{0.10}_{0.14} $ &      $  126.1 \pm   5.1 $ &  $	 13.89 \pm\,^{0.08}_{0.10} $ &  	 \nodata	 &  $ <  18.58  	   $ &      $	  0.20 \pm\,^{0.13}_{0.18} $ &      $ >    -0.72		   $ \\
  Sk--67\D191     & $  5.5595 $  & $-67.5054 $ &    $  131.9 \pm   2.4 $ &  $	 15.21 \pm 0.03 	   $ &      $  129.3 \pm   0.8 $ &  $	 14.03 \pm 0.02 	   $ &      $  131.7 \pm   1.5 $ &  $	 18.77 \pm 0.12 $ &	    $	  0.09 \pm 0.04 	   $ &      $	 -0.29 \pm 0.12 	   $ \\
       HV5936     & $  5.5608 $  & $-66.6277 $ &	 \nodata	 &  $ <  14.47  		   $ &  	 \nodata	 &  $ <  13.26  		   $ &  	 \nodata	 &  $ <  18.47  	   $ &  	  \nodata		     &  	  \nodata		     \\
        BI208     & $  5.5660 $  & $-67.4056 $ &    $  114.8 \pm   1.9 $ &  $	 14.50 \pm\,^{0.06}_{0.07} $ &      $  115.8 \pm   1.4 $ &  $	 13.69 \pm 0.04 	   $ &  	 \nodata	 &  $ <  18.47  	   $ &      $	  0.46 \pm\,^{0.07}_{0.09} $ &      $ >    -0.94		   $ \\
  Sk--69\D191     & $  5.5721 $  & $-69.7528 $ &    $  110.9 \pm   1.2 $ &  $	 15.09 \pm 0.04 	   $ &      $  110.0 \pm   1.7 $ &  $	 13.94 \pm\,^{0.06}_{0.07} $ &  	 \nodata	 &  $ <  18.46  	   $ &      $	  0.12 \pm\,^{0.07}_{0.09} $ &      $ >    -0.29		   $ \\
 J053441-6931     & $  5.5781 $  & $-69.5275 $ &	 \nodata	 &     \nodata  		     &      $  111.4 \pm   3.3 $ &  $	 13.41 \pm\,^{0.10}_{0.12} $ &  	 \nodata	 &  $ <  18.42  	   $ &  	  \nodata		     &  	  \nodata		     \\
  Sk--67\D205     & $  5.5814 $  & $-67.2724 $ &    $  108.9 \pm   3.1 $ &  $	 14.82 \pm\,^{0.07}_{0.09} $ &      $  101.4 \pm   1.7 $ &  $	 14.06 \pm 0.04 	   $ &  	 \nodata	 &  $ <  18.56  	   $ &      $	  0.51 \pm\,^{0.08}_{0.10} $ &      $ >    -0.74		   $ \\
  Sk--67\D211$^a$ & $  5.5872 $  & $-67.5575 $ &    $  111.8 \pm   0.0 $ &  $	 14.79 \pm 0.05 	   $ &      $  112.4 \pm   1.6 $ &  $	 13.98 \pm 0.02 	   $ &  	 \nodata	 &  $ <  18.53  	   $ &      $	  0.46 \pm 0.03 	   $ &      $ >    -0.67		   $ \\
        BI237     & $  5.6041 $  & $-67.6554 $ &	 \nodata	 &     \nodata  		     &      $  109.6 \pm   3.0 $ &  $	 14.01 \pm 0.05 	   $ &  	 \nodata	 &  $ <  18.62  	   $ &  	  \nodata		     &  	  \nodata		     \\
  Sk--66\D169     & $  5.6151 $  & $-66.6403 $ &	 \nodata	 &  $ <  14.07  		   $ &  	 \nodata	 &  $ <  13.13  		   $ &  	 \nodata	 &  $ <  18.47  	   $ &  	  \nodata		     &  	  \nodata		     \\
  Sk--69\D223     & $  5.6156 $  & $-69.1933 $ &	 \nodata	 &     \nodata  		     &  	 \nodata	 &  $ <  13.26  		   $ &  	 \nodata	 &  $ <  18.47  	   $ &  	  \nodata		     &  	  \nodata		     \\
  Sk--66\D171     & $  5.6173 $  & $-66.6442 $ &	 \nodata	 &  $ <  14.43  		   $ &  	 \nodata	 &  $ <  13.37  		   $ &  	 \nodata	 &  $ <  18.47  	   $ &  	  \nodata		     &  	  \nodata		     \\
  Sk--66\D172     & $  5.6182 $  & $-66.3599 $ &	 \nodata	 &     \nodata  		     &  	 \nodata	 &  $ <  13.35  		   $ &  	 \nodata	 &  $ <  18.47  	   $ &  	  \nodata		     &  	  \nodata		     \\
        BI253     & $  5.6262 $  & $-69.0194 $ &	 \nodata	 &     \nodata  		     &  	 \nodata	 &  $ <  13.36  		   $ &  	 \nodata	 &  $ <  18.47  	   $ &  	  \nodata		     &  	  \nodata		     \\
  Sk--68\D135     & $  5.6302 $  & $-68.9189 $ &	 \nodata	 &     \nodata  		     &      $  132.1 \pm   1.6 $ &  $	 14.20 \pm 0.04 	   $ &  	 \nodata	 &  $ <  18.60  	   $ &  	  \nodata		     &  	  \nodata		     \\
  Sk--68\D137     & $  5.6402 $  & $-68.8758 $ &	 \nodata	 &     \nodata  		     &  	 \nodata	 &  $ <  13.38  		   $ &  	 \nodata	 &  $ <  18.47  	   $ &  	  \nodata		     &  	  \nodata		     \\
  Sk--69\D243     & $  5.6452 $  & $-69.1009 $ &	 \nodata	 &     \nodata  		     &  	 \nodata	 &  $ <  13.33  		   $ &  	 \nodata	 &  $ <  18.47  	   $ &  	  \nodata		     &  	  \nodata		     \\
  Sk--69\D246$^a$ & $  5.6482 $  & $-69.0335 $ &    $  115.6 \pm   1.0 $ &  $	 14.21 \pm 0.02 	   $ &      $  115.1 \pm   3.7 $ &  $	 13.24 \pm\,^{0.11}_{0.17} $ &  	 \nodata	 &  $ <  18.58  	   $ &      $	  0.30 \pm\,^{0.11}_{0.17} $ &      $ >    -1.24		   $ \\
  Sk--69\D249     & $  5.6495 $  & $-69.4886 $ &	 \nodata	 &     \nodata  		     &      $  105.9 \pm   2.0 $ &  $	 13.47 \pm 0.05 	   $ &  	 \nodata	 &  $ <  18.44  	   $ &  	  \nodata		     &  	  \nodata		     \\
  Sk--69\D257     & $  5.6664 $  & $-69.7342 $ &    $  108.7 \pm   3.4 $ &  $	 14.60 \pm\,^{0.09}_{0.11} $ &      $	94.5 \pm   3.5 $ &  $	 13.51 \pm\,^{0.11}_{0.14} $ &      $  105.9 \pm   5.0 $ &  $	 18.52 \pm 0.16 $ &	    $	  0.18 \pm\,^{0.13}_{0.19} $ &      $	 -0.65 \pm\,^{0.17}_{0.21} $ \\
  Sk--66\D185     & $  5.7085 $  & $-66.3030 $ &	 \nodata	 &  $ <  14.39  		   $ &  	 \nodata	 &     \nodata  		     &  	 \nodata	 &  $ <  18.47  	   $ &  	  \nodata		     &  	  \nodata		     \\
    D301-1005     & $  5.7190 $  & $-67.8479 $ &	 \nodata	 &  $ <  14.36  		   $ &  	 \nodata	 &  $ <  13.37  		   $ &  	 \nodata	 &  $ <  18.47  	   $ &  	  \nodata		     &  	  \nodata		     \\
      LH114-7     & $  5.7202 $  & $-67.8545 $ &	 \nodata	 &  $ <  14.14  		   $ &      $  152.2 \pm   4.7 $ &  $	 13.42 \pm\,^{0.10}_{0.12} $ &  	 \nodata	 &  $ <  18.51  	   $ &      $ >    0.43 		   $ &  	  \nodata		     \\
  Sk--67\D250     & $  5.7210 $  & $-67.8527 $ &	 \nodata	 &     \nodata  		     &      $	92.3 \pm   2.8 $ &  $	 13.67 \pm\,^{0.07}_{0.09} $ &  	 \nodata	 &  $ <  18.41  	   $ &  	  \nodata		     &  	  \nodata		     \\
     D301-NW8     & $  5.7211 $  & $-67.8308 $ &	 \nodata	 &  $ <  14.38  		   $ &      $  165.6 \pm   4.2 $ &  $	 13.34 \pm\,^{0.13}_{0.19} $ &  	 \nodata	 &  $ <  18.36  	   $ &      $ >    0.04 		   $ &  	  \nodata		     \\
        BI272     & $  5.7398 $  & $-67.2415 $ &    $  144.7 \pm   1.7 $ &  $	 14.95 \pm 0.03 	   $ &      $  138.6 \pm   1.4 $ &  $	 13.67 \pm 0.04 	   $ &      $  141.0 \pm   6.0 $ &  $	 18.64 \pm 0.13 $ &	    $	 -0.01 \pm 0.05 	   $ &      $	 -0.42 \pm 0.13 	   $ \\
  Sk--67\D266     & $  5.7644 $  & $-67.2403 $ &	 \nodata	 &  $ <  14.31  		   $ &  	 \nodata	 &  $ <  13.25  		   $ &  	 \nodata	 &  $ <  18.47  	   $ &  	  \nodata		     &  	  \nodata		     \\
  Sk--70\D115     & $  5.8138 $  & $-70.0660 $ &	 \nodata	 &     \nodata  		     &  	 \nodata	 &  $ <  13.01  		   $ &  	 \nodata	 &  $ <  18.47  	   $ &  	  \nodata		     &  	  \nodata		     \\
  Sk--68\D171     & $  5.8397 $  & $-68.1907 $ &	 \nodata	 &  $ <  14.46  		   $ &  	 \nodata	 &     \nodata  		     &  	 \nodata	 &  $ <  18.47  	   $ &  	  \nodata		     &  	  \nodata		     \\
  Sk--70\D120     & $  5.8558 $  & $-70.2858 $ &	 \nodata	 &     \nodata  		     &  	 \nodata	 &  $ <  13.36  		   $ &  	 \nodata	 &  $ <  18.47  	   $ &  	  \nodata		     &  	  \nodata		     
\enddata
\tablecomments{$a$: Stars that have spectra from STIS E140M and \fuse. Measurements and limits on $N($\oi$)$ and N$($\feii$)$ are primarily based on the \oi\ $\lambda$1039 and \feii\ $\lambda$1144 lines, except for Sk--69\D246 where  \oi\ $\lambda$1302 and \feii\ $\lambda$1608 only  are used as the weaker transitions are not detected. The \fuse\ velocities (\oi\ and \feii) have an additional systematic error of about 5--10 \km. For $[$O/H$]$, there is an additional $\pm 0.08$ dex ``beam'' error (see text for more details). For the limits on the $[$O/H$]$, the errors on $N($\oi$)$ or $N($\hi$)$ were taken into account, i.e. the upper limits are quoted as the limit $+ 1\sigma$ (statistical uncertainty). The adopted solar references are $[$O/H$]_\odot = -3.27$ and $[$Fe/O$]_\odot = -1.26$ from \citet{lodders09}.}
\end{deluxetable}
\clearpage
\end{landscape}

\begin{deluxetable}{lccc}
\tablewidth{0pc}
\tablecaption{Relative Abundances in the HVC  \label{t-rel}}
\tabletypesize{\footnotesize}
\tablehead{\colhead{Name} & \colhead{$[$\siii/\oi$]$}& \colhead{$[$\siii/\sii$]$} & \colhead{$[$\feii/\siii$]$}}
\startdata
    4U0532-664   	& $ +0.68 \pm 0.03 		$   & $>+0.23	    	$  & $ -0.06 \pm 0.04      	$   \\
    BRRG140      	& $ +0.33 \pm 0.07 		$   & $>-0.48	    	$  & $ -0.08 \pm 0.09      	$   \\
    Brey22$^a$       	& $ +0.37 \pm 0.06 		$   & $ -0.22 \pm 0.10  $  & $ -0.03 \pm 0.06	 	$   \\
    LH54-425$^a$     	& $ +0.32 \pm 0.06		$   & $ -0.28 \pm 0.12  $  & $ -0.18 \pm 0.04	 	$   \\
    Sk--65\D22   	& $ +0.70 \pm \,^{0.16}_{0.12}	$   & $ >-0.12	    	$  & $ -0.14 \pm 0.08      	$ \\
    Sk--67\D101  	& $ +0.47 \pm 0.06 		$   & $ +0.03 \pm 0.12 	$  & $ -0.18 \pm 0.06	 	$   \\
    Sk--67\D104  	& $>+0.33         		$   & $ >-0.05	    	$  & $>-0.03	 		$   \\
    Sk--67\D106 	& $ +0.23 \pm 0.08 		$   & $ -0.15 \pm 0.12 	$  & $ -0.06 \pm 0.08	 	$	    \\
    Sk--67\D107  	& $ +0.38 \pm 0.06 		$   & $ -0.10 \pm 0.14 	$  & $ -0.04 \pm 0.04	 	$   \\
    Sk--67\D111$^a$  	& $ +0.34 \pm 0.06		$   & $ -0.11 \pm 0.09  $  & $ -0.06 \pm 0.04	 	$   \\
    Sk--67\D166$^a$  	& $ +0.38 \pm 0.06		$   & $ >-0.08 \pm 0.14 $  & $ -0.03 \pm 0.03	 	$   \\
    Sk--67\D211  	& $ +0.50 \pm 0.05 		$   & $> -0.44	    	$  & $ -0.04 \pm 0.03      	$   \\
    Sk--69\D246  	& $ +0.60 \pm 0.03 		$   & $> -0.56	    	$  & $ >-0.30			$	\\
    Sk--71\D45   	& $ +0.20 \pm 0.03 		$   & $ -0.25 \pm 0.10 	$  & $ -0.09 \pm 0.03	 	$   
\enddata
\tablecomments{All stars are from the STIS+\fuse\ sample except for those marked with $a$ that are from \fuse\ only  and where \pii\ is taken as a proxy for \sii. The adopted solar references are $[$Si/O$]_\odot = -1.20$, $[$Si/S$]_\odot = +0.37$, $[$Si/P$]_\odot = +2.08$, and $[$Fe/Si$]_\odot = -0.07$ from \citet{lodders09}.
}
\end{deluxetable}

\clearpage

\end{document}